\documentclass[superscriptaddress,twocolumn,floatfix,prb,aps]{revtex4-1}
\usepackage{graphicx}
\usepackage{amsmath}
\usepackage{natbib}
\usepackage[pdftex,breaklinks,bookmarks=false,colorlinks,linkcolor=blue,citecolor=blue,urlcolor=blue]{hyperref}

\begin{document}


\title{Equivalent critical behavior of a helical point contact and a two-channel Luttinger liquid - topological superconductor junction}

\author{C. L. Kane}
\affiliation{ Department of Physics and Astronomy, University of Pennsylvania, Philadelphia, PA 19104-6323, USA}
\author{D. Giuliano}
\affiliation{Dipartimento di Fisica, Universit\`a della Calabria, Arcavacata di Rende, I-87036, Cosenza, Italy}
\affiliation{I.N.F.N., Gruppo Collegato di Cosenza, Arcavacata di Rende, I-87036, Cosenza, Italy}
\author{I. Affleck}
\affiliation{Department of Physics and Astronomy and Stewart Blusson Quantum Matter Institute,
University of British Columbia, Vancouver, B.C., Canada, V6T 1Z1}

\begin{abstract}
We demonstrate the equivalence between two distinct Luttinger liquid impurity problems.   The first concerns a one-dimensional topological superconductor 
coupled at one end to the ends of two single channel Luttinger liquids.   The second concerns a point contact in the quantum spin Hall effect, where four helical
Luttinger liquids meet at a point.   Both of these problems have been studied previously, and exhibit several stable phases, depending on the Luttinger parameter $K$,
that can be characterized in terms of simple conformally invariant boundary conditions describing  perfect normal (or Andreev) transmission or reflection.   In addition, 
both problems exhibit critical points that are described by ``intermediate" fixed points similar to those found in earlier studies of an impurity in a Luttinger liquid with spin.  
Though these two models have different symmetries and numbers of modes, we show they are equivalent and are related by a duality transformation, and we show that the
non-trivial intermediate critical points are the same.  In the non-interacting limit, $K=1$, the duality involves two distinct free fermion representations that are related 
by a non-local transformation that derives from the triality of $SO(8)$.   Using the explicit translation between the two theories, we translate results from one problem to 
the other and vice versa.   This allows us to make new predictions about the topological superconductor-Luttinger liquid junction, including predictions about the global 
behavior of the critical conductance $G^*(K)$, as well predictions for the critical exponents and universal crossover scaling functions.
In this paper we introduce both models from scratch, using a common notation that facilitates their comparison, and we discuss in detail the dualities that relate them,
along with their free fermion limits.   We close with a discussion of open problems and future directions.

\end{abstract}

\maketitle

\section{Introduction}

Quantum impurity problems have played a central role in the development of quantum many body theory.  A central paradigm, introduced by Affleck and Ludwig\cite{affleck91a,affleck91b}, is that the fixed points characterizing the low energy phases of a 0+1 dimensional impurity coupled to a bath are in correspondence with the allowed conformally invariant boundary conditions of the conformal field theory describing the bath.   Applying the powerful techniques of boundary conformal field theory\cite{cardy84}, this allows for a detailed characterization of non-Fermi liquid behavior that arises in the multi-channel Kondo problem\cite{affleck91a,affleck91b}, the single impurity problem in a Luttinger liquid\cite{kanefisher92a,kanefisher92b,furusaki93}, the theory of point contacts in the fractional quantum Hall effect\cite{moon93,fendley95}, and many related problems.

In the Luttinger liquid problem, the simplest boundary conditions for a weak link are the ``perfectly transmitting" (``perfectly reflecting") limits, which are stable for attractive (repulsive) interactions.   In a Luttinger liquid with spin there are additional fixed points in which charge is perfectly transmitted and spin is perfectly reflected, or vice versa\cite{kanefisher92b,furusaki93}.   The CI, IC, CC and II (charge conductor, spin insulator, etc.) phases are stable or unstable, depending on the values of the Luttinger parameters $K_\rho$ and $K_\sigma$ characterizing the interactions, and are described by simple boundary conditions in the charge and spin sectors.   However, for certain ranges of $K_\rho$, $K_\sigma$ it was found that all of the ``simple" fixed points are unstable, or that more than one is stable.   This implies the existence of non-trivial additional fixed points.  
A perturbative analysis of these fixed points is possible in limits where they are close to the simpler fixed points\cite{kanefisher92b}, and for certain specific values of the interactions the intermediate fixed points can be mapped to solvable models\cite{yikane98,affleck01}.   However, a complete theory of these ``intermediate" fixed points has remained elusive.

In the problem of the spinful Luttinger liquid, the intermediate fixed points arise in a rather unphysical parameter regime, $K_\rho<1$ and $K_\sigma >1$.   However, Hou, Kim and Chamon\cite{hou09} pointed out that a point contact in a quantum spin Hall insulator maps to a weak link in a spinful Luttinger liquid with Luttinger parameters $K_\rho = 1/K_\sigma = K$, where $K$ is the Luttinger parameter characterizing the helical edge state of the quantum spin Hall insulator, which forms a non-chiral Luttinger liquid\cite{wu06,xu06}.   This led Teo and Kane\cite{teokane09} (TK) to develop a theory of the critical behavior of the pinch-off transition of a helical point contact.   
For the helical point contact, both the pinched off and the open limits (which both correspond to simple conformally invariant boundary conditions) are perturbatively stable when $1/2<K<2$.   In both cases, the perturbative corrections involve tunneling of electrons between the middles of two Luttinger liquids, which is irrelevant for any $K \ne 1$.   For $1/2<K<2$, the pinchoff transition is controlled by an intermediate unstable fixed point.   At zero temperature, the conductance changes discontinuously at the transition, while at finite temperature, the transition has a finite width and is described by a universal crossover scaling function.
For $K < 1/2$ ($K > 2$), the system flows to the charge insulator - spin conductor (charge conductor - spin insulator) fixed point with zero (perfect) conductance.
By piecing together perturbative solutions at $K = 1/2 + \epsilon$, $K = 2 - \epsilon$ and $1 \pm \epsilon$, along with symmetry arguments at $K = 1/\sqrt{3}$ and $K = \sqrt{3}$, and assuming there are no additional fixed points, TK predicted the behavior of the intermediate critical point as a function of $K$.

In a subsequent, but independent development, Affleck and Giuliano\cite{affleck13,giuliano19} (AG) studied the problem of a junction between Luttinger liquids and a topological superconductor.   This was motivated by the proposal to realize one dimensional topological superconductivity in nanowires proximitized by a superconductor\cite{lutchyn10,oreg10,mourik12}.   When the Majorana zero mode at the end of a 1D topological superconductor is weakly coupled to a Fermi liquid lead, then at high energy the electrons are normally reflected, but in the limit of low energy the electrons exhibit perfect Andreev reflection\cite{law09}.   At low energy, the Majorana mode is effectively absorbed by the lead, resulting in a change in the boundary condition from normal to Andreev reflection.    A similar phenomenon occurs when a one dimensional Luttinger liquid is coupled to the Majorana mode, provided the interactions are not too large\cite{fidkowski12}.   For $K > 1/2$ the Majorana mode is absorbed, resulting in Andreev reflection, while for $K<1/2$ there is Normal reflection, with a decoupled Majorana mode.

AG considered the case in which the Majorana mode is coupled to two Luttinger liquid leads.   When $K<1/2$ the Majorana mode is decoupled from both leads, which both have Normal reflection (the NN phase).   For $K>1/2$, when lead 1 is more strongly coupled than lead 2, then at low energy the Majorana mode is absorbed by lead 1, leading to Andreev (Normal) reflection in lead 1 (2)  (the AN phase).    When the couplings to leads 1 and 2 are equal, however, the Majorana mode is frustrated.   AG showed that the low energy behavior in this case is controlled by a non-trivial intermediate fixed point, which can be described perturbatively for $K = 1/2 + \epsilon$ and for free fermions when $K=1$.   

In this paper we show that the helical point contact described by the TK model and the Luttinger liquid-topological superconductor junction, described by the AG model are equivalent and related by a duality transformation.   The (II, CC, IC, CI) phases of the TK model correspond to (NA, AN, NN, AA) phases of the AG model, and the intermediate fixed point that describes the pinch off transition is the same as the fixed point that characterizes the symmetric junction.   The mapping between the models allows us to transfer many of the results from TK to the AG model.   Specifically, we show that the critical conductance of the AG model exhibits a symmetry under $K\leftrightarrow K^{-1}$, and we adopt the perturbative results of TK for $K=1+\epsilon$, as well as the behavior at $K = 1/\sqrt{3}$ and $K=\sqrt{3}$ to form a more complete picture of the behavior of the conductance as a function of $K$.  In addition, we translate the results of TK for the critical exponents and universal crossover scaling functions to the AG model.

In section \ref{section2} we will introduce the TK model of the helical point contact and describe the web of dualities that provide equivalent descriptions of the model and are useful for describing the different phases.   Much of the material in this section is contained in TK.   We include it here to introduce notation that facilitates comparison with AG.   In section \ref{section3}, we introduce the AG model, and in section \ref{section4} we demonstrate the equivalence with the TK model by explicitly describing the duality that relates them.   In addition, we consider the free fermion limit $K=1$ and show that the two models correspond to two inequivalent free fermion representations that are related by a transformation that derives from the triality of $SO(8)$.
Finally, in section \ref{section5} we describe new predictions about the AG model that follow from our identification before concluding in section \ref{section6}.

\section{Helical Point Contact}
\label{section2}

A point contact in the quantum spin Hall effect involves four helical edges, labeled by $a = 1, ..., 4$, that meet at a point.  Each helical edge consists of counterpropagating Dirac fermions modes, described by $\psi_{a,p}(x)$, where $p = {\rm out}, {\rm in}$ (also denoted by $p= +1$, $-1$) specifies the direction of propagation, and $x$ is the distance from the contact.   In the spin $S_z$ conserving model that we consider, the spin of each mode is correlated with its propagation direction and given by $S_z = p(-1)^a \hbar/2$.

\begin{figure}
\includegraphics[width=3.4in]{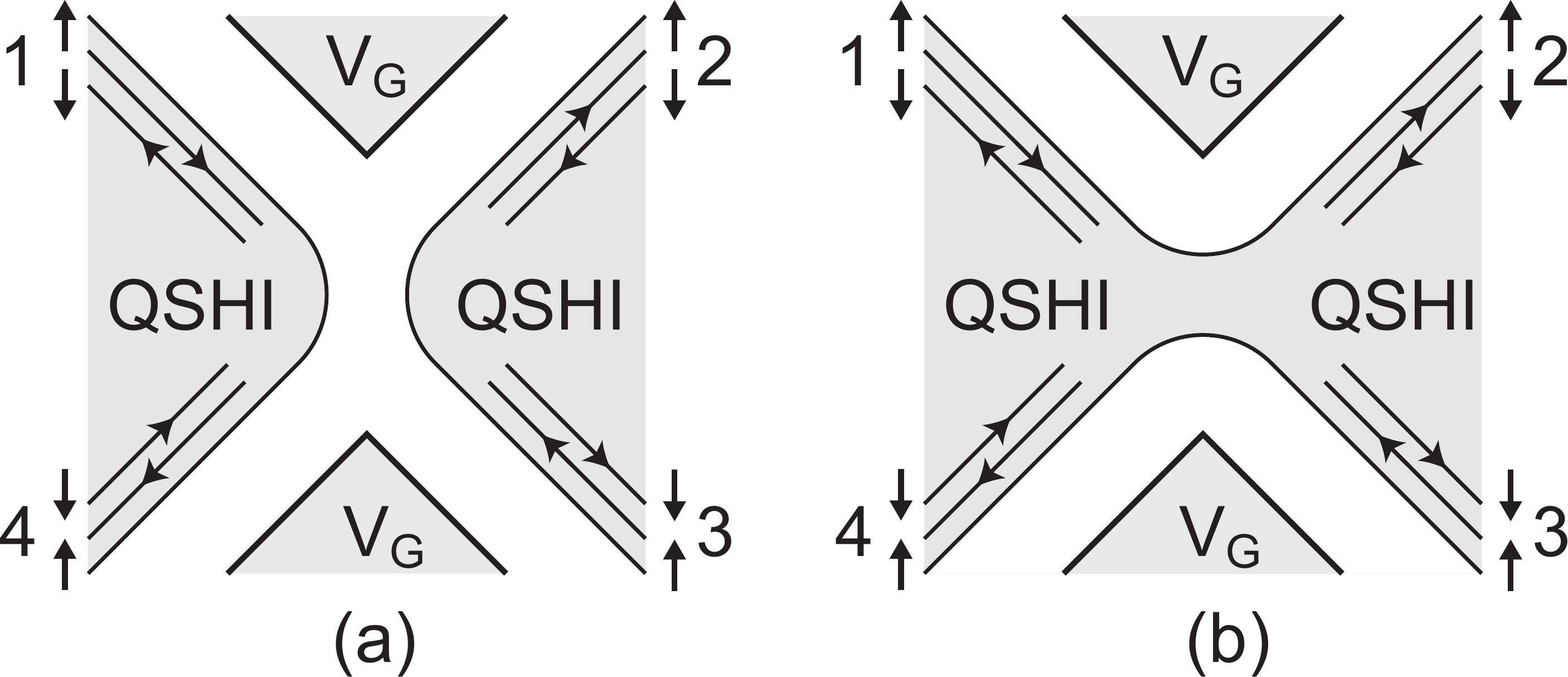}
\caption{A point contact in a quantum spin Hall insulator in the (a) pinched off limit and (b) the open limit. }
\label{Fig1}
\end{figure}

The point contact admits a simple description in the pinched-off and open limits, shown in Fig. \ref{Fig1}(a) and \ref{Fig1}(b).   These lead to the simple boundary conditions on the chiral fermion modes.   For the pinched off junction, we have
\begin{align}
\psi_{1,{\rm out}}(0) &= \psi_{4,{\rm in}}(0); \quad
\psi_{2,{\rm out}}(0) = \psi_{3,{\rm in}}(0); \nonumber\\
\psi_{3,{\rm out}}(0) &= \psi_{2,{\rm in}}(0); \quad
\psi_{4,{\rm out}}(0) = \psi_{1,{\rm in}}(0),
\label{psibcii}
\end{align}
while for the open junction we have
\begin{align}
\psi_{1,{\rm out}}(0) &= \psi_{2,{\rm in}}(0); \quad
\psi_{2,{\rm out}}(0) = \psi_{1,{\rm in}}(0); \nonumber\\
\psi_{3,{\rm out}}(0) &= \psi_{4,{\rm in}}(0); \quad
\psi_{4,{\rm out}}(0) = \psi_{3,{\rm in}}(0).
\label{psibccc}
\end{align}
For non-interacting fermions, a more general boundary condition can be expressed in terms a unitary $4\times 4$ transmission matrix.   We will discuss this in Section \ref{freefermionsec}.   Here we focus on the interacting case, where the helical edge states form a Luttinger liquid.   The Luttinger liquid theory can be formulated by expanding perturbatively about either of the above limits.

\subsection{Luttinger liquid model}

\subsubsection{Four-channel variables}

In the presence of short ranged electron interactions, each of the four helical edges forms a non-chiral Luttinger liquid.  This is most easily described by bosonizing
\begin{equation}
\psi_{a,p} = \frac{e^{i (\varphi_a + p \theta_a)}}{\sqrt{2\pi x_c}} ,
\label{psibosonize}
\end{equation}
where $x_c$ is a short distance cut-off, and the boson fields satisfy $[\theta_a(x),\theta_{a'}(x')]=[\varphi_a(x),\varphi_{a'}(x')] = 0$ along with 
\begin{equation}
[\partial_x\theta_a(x), \varphi_{a'}(x') ] = i\pi \delta_{aa'}\delta(x-x').
\label{phicommute}
\end{equation}
The boundary condition at $x=0$ is determined by (\ref{psibcii}) or (\ref{psibccc}).

In general, (\ref{psibosonize},\ref{phicommute}) should be augmented with either a Klein factor in (\ref{psibosonize}) or a specification of the commutator of the zero modes of $\theta_{a}$ and $\varphi_{a}$ that ensures that the Fermi fields $\psi_{a,p}$ all anticommute with one another.  For the problem at hand, which will involve tunneling of electrons between the helical edges, while conserving charge and spin, the zero modes and the Klein factors have no effect and can be ignored.

In the presence of interactions, the Hamiltonian has the form,
\begin{equation}
H = \int_0^\infty dx \sum_{a = 1}^4 {\cal H}^0_a(x),
\end{equation}
where
\begin{equation}
{\cal H}^0_a = \frac{v}{2\pi} \left[ K^{-1} (\partial_x\theta_a)^2 + K(\partial_x\varphi_a)^2 \right].
\label{h0a}
\end{equation}
$K$ is the dimensionless Luttinger parameter characterizing the forward scattering interactions and $v$ is the velocity.   $K <1$ ($K>1$) for repulsive (attractive) interactions.   For the 
operators built from products of electron operators
 $\psi_{a,mn} = \exp i( m \varphi_a + n\theta_a)$ (for $n=m$ mod 2), $K$ determines the scaling dimension
\begin{equation}
\Delta_{mn} = (K^{-1} m^2 + K n^2)/4.
\label{dimensionn1n2}
\end{equation}

The densities of the conserved charge and spin in lead $a$ are $n^\rho_a = \partial_x\theta_a/\pi$ and $n^\sigma_a = (-1)^{a-1} \partial_x\varphi_a/\pi$.   It follows that the electric current flowing into lead $a$ is 
\begin{equation}
I^\rho_a = \partial_t \theta_a/\pi = v K \partial_x \varphi_a/\pi.
\end{equation}
and the spin current into lead $a$ is
\begin{equation}
I^\sigma_a = (-1)^{a-1} \partial_t \varphi_a/\pi = v (-1)^{a-1} K^{-1} \partial_x \theta_a/\pi
\end{equation}
Of interest are the currents of charge and spin flowing from left to right (from leads 1 and 4 to leads 2 and 3)
\begin{align}
I^\rho_X &= (I^\rho_1 - I^\rho_2 - I^\rho_3 + I^\rho_4)/2 \\
I^\sigma_X &= (I^\sigma_1 - I^\sigma_2 - I^\sigma_3 + I^\sigma_4)/2 
\end{align}
and the currents flowing from top to bottom (from leads 1 and 2 to leads 3 and 4)
\begin{align}
I^\rho_Y &= (I^\rho_1 + I^\rho_2 - I^\rho_3 - I^\rho_4)/2 \\
I^\sigma_Y &= (I^\sigma_1 + I^\sigma_2 - I^\sigma_3 - I^\sigma_4)/2 .
\end{align}
These define charge and spin conductances, computed by the Kubo formula as  $G_{IJ}^\alpha= {\rm lim}_{\omega\rightarrow 0} \Pi_{IJ}^\alpha(\omega)/(i\omega)$, for $I,J = X$ or $Y$,
where the retarded current-current correlation function is
\begin{equation}
\Pi_{IJ}^\alpha(t) = \theta(t) \langle [I_I^\alpha(t), I_J^\alpha(0)]\rangle.
\label{kuboformula}
\end{equation}

\subsubsection{Charge and Spin Variables}

Due to the conservation of charge and spin at the point contact it is useful to introduce new variables,
\begin{align}
\varphi_{\rho \pm} &= (\varphi_1 \pm \varphi_2 \pm \varphi_3 + \varphi_4)/2,\label{vv1} \\
\theta_{\rho \pm} &= (\theta_1 \pm \theta_2 \pm \theta_3 + \theta_4)/2, 
\label{vv2}\end{align}
and
\begin{align}
\varphi_{\sigma \pm} &= (\theta_1 \mp \theta_2 \pm \theta_3 - \theta_4)/2, \label{vv3}\\
\theta_{\sigma \pm} &= (\varphi_1 \mp \varphi_2 \pm \varphi_3 - \varphi_4)/2.
\label{vv4}
\end{align}

These variables satisfy $[\theta_\alpha(x),\theta_{\alpha'}(x')]=[\varphi_\alpha(x),\varphi_{\alpha'}(x')] = 0$ for $\alpha,\alpha' = \rho\pm, \sigma\pm$, along with 
\begin{equation}
[\partial_x\theta_\alpha(x), \varphi_{\alpha'}(x') ] = i\pi \delta_{\alpha\alpha'}\delta(x-x').
\label{thetaalphacommute}
\end{equation}
In terms of these variables the Hamiltonian is
\begin{equation}
H = \int_0^\infty dx \left[ {\cal H}^0_{\rho +} + {\cal H}^0_{\sigma  +} + {\cal H}^0_{\rho-} + {\cal H}^0_{\sigma -}\right]
\label{hh}
\end{equation}
with
\begin{align}
{\cal H}^0_{\rho\pm} &= \frac{v}{2\pi} \left[ K_\rho^{-1}(\partial_x\theta_{\rho\pm})^2 + K_\rho(\partial_x\varphi_{\rho\pm})^2\right] \label{hrho0}\\
{\cal H}^0_{\sigma\pm} &= \frac{v}{2\pi} \left[K_\sigma^{-1} (\partial_x\theta_{\sigma\pm})^2 + K_\sigma(\partial_x\varphi_{\sigma\pm})^2\right], \label{hsigma0}
\end{align}
and
\begin{equation}
K_\rho = K; \quad K_\sigma = K^{-1}.  \label{krhosigma}
\end{equation}

In terms of these variables, we have 
\begin{align}
I^\rho_X &= \partial_t\theta_{\rho -}/\pi = K v\partial_x\varphi_{\rho-}/\pi \label{irhox}\\
I^\sigma_X &= \partial_t\theta_{\sigma -}/\pi = K^{-1} v \partial_x\varphi_{\sigma-}/\pi \label{isigmax}
\end{align}
along with
\begin{align}
I^\rho_Y &= \partial_t\varphi_{\sigma -}/\pi = K v \partial_x\theta_{\sigma -}/\pi \label{irhoy}\\
I^\sigma_Y &= \partial_t\varphi_{\rho -}/\pi = K^{-1}  v\partial_x\theta_{\rho-}/\pi \label{isigmay}
\end{align}
The total charge and spin flowing out of the junction are
\begin{align}
\sum_{a=1}^4 I^\rho_a = 2\partial_t \theta_{\rho+}/\pi = 2 v K \partial_x \varphi_{\rho+}/\pi,\\
\sum_{a=1}^4 I^\sigma_a = 2\partial_t \theta_{\sigma+}/\pi = 2 v K^{-1} \partial_x \varphi_{\sigma+}/\pi.
\end{align}
Charge and spin conservation implies that these operators have zero expectation values.

\subsection{Boundary Conditions and Fixed Points}

Note that the currents depend only on the variables in the $\rho-$ and $\sigma-$ sectors.    The boundary conditions (\ref{psibcii}) or (\ref{psibccc}) imply that the boundary condition in the $\rho +$ and $\sigma + $ sectors is
\begin{equation}
\theta_{\rho +}(0) = 0; \quad \theta_{\sigma +}(0) = 0.
\end{equation}
Since charge and spin conservation at the junction forbids terms in the Hamiltonian proportional to $\exp i \varphi_{\rho +}$ or $\exp i \varphi_{\sigma +}$, the $\rho +$ and $\sigma +$ sectors are trivially perfectly transmitted and decouple from $\rho -$ and $\sigma -$ sectors.  The non-trivial behavior of the point contact is all in the $\rho -$ and $\sigma -$ sectors, which we describe perturbatively below.

\subsubsection{Closed junction:  II fixed point:}  

For a closed junction, the boundary condition (\ref{psibcii}) implies that
\begin{equation}
\theta_{\rho -}(0) = 0; \quad \theta_{\sigma -}(0) = 0.
\label{iibc}
\end{equation}
This corresponds to $G_{XX}^\rho = G_{XX}^\sigma = 0$, which we refer to as the charge insulator/spin insulator (II) phase.    Given (\ref{iibc}) we can integrate out the degrees of freedom for $x>0$ in an imaginary time path integral to obtain an effective 0+1D action for $\varphi_{\alpha}(\tau) \equiv \varphi_{\alpha}(x=0,\tau)$, given by
\begin{equation}
S_0^{II} =  \int \frac{d\omega}{(2\pi)^2}\left[ K_\rho |\omega||\varphi_{\rho-}(\omega)|^2 + 
K_\sigma |\omega||\varphi_{\sigma-}(\omega)|^2\right]
\label{s0ii}
\end{equation}
where $\omega$ is a Matsubara frequency and $K_{\rho,\sigma}$ are given in (\ref{krhosigma}).

The transfer of electrons across the junction can be described perturbatively in terms of the tunneling Hamiltonian
\begin{equation}
V_{II} = -t_e \cos \varphi_{\rho-} \cos \varphi_{\sigma-} - t_\rho \cos 2\varphi_{\rho-} - t_{\sigma-} \cos 2 \varphi_{\sigma-},
\label{Vt}
\end{equation}
Here $t_e$ describes the tunneling of electrons (with $\uparrow$ or $\downarrow$)
across the junction and $t_\rho$ describes the tunneling of a 
$\uparrow\downarrow$ pair of electrons.   $t_\sigma$ describes the transmission of spin by tunneling of a $\uparrow$($\downarrow$) electron to the right (left).    In general, one could also consider higher order tunneling processes, but they will be less relevant.   The stability of the II fixed point is determined by the RG flow
equation $dt/d\ell = (1-\Delta(t))t$, where from (\ref{dimensionn1n2}) the scaling dimensions for the tunneling
operators are given by
\begin{equation}
\Delta(t_e) = (K + K^{-1})/2; \enskip \Delta(t_\rho) = 2 K^{-1}; \enskip \Delta(t_\sigma) = 2 K.
\label{dimensiont}
\end{equation}
For $1/2<K<2$, all perturbations are irrelevant, and the II fixed point is stable.   For $K<1/2$ ($K>2$), $t_\sigma$ ($t_\rho$) is relevant and flows to strong coupling.

\subsubsection{Open junction: CC fixed point}

When the junction is open, the boundary condition (\ref{psibccc}) implies
\begin{equation}
\varphi_{\rho -}(0) = 0; \quad \varphi_{\sigma -}(0) = 0.
\label{ccbc}
\end{equation}
In this case the transmission of charge and spin is perfect,  $G_{XX}^\rho = G_{XX}^\sigma = 2 K e^2/h$, which we call the charge conductor/spin conductor (CC) phase.
We then have
\begin{equation}
S_0^{CC} =  \int \frac{d\omega}{(2\pi)^2}\left[ K_\rho^{-1} |\omega||\theta_{\rho-}(\omega)|^2 + 
K_\sigma^{-1} |\omega||\theta_{\sigma-}(\omega)|^2\right]
\label{s0cc}
\end{equation}

The perturbative corrections to this fixed point involve tunneling electrons between the top and bottom edges, and are described by the tunneling Hamiltonian
\begin{equation}
V_{CC} = v_e \cos \theta_{\rho-} \cos \theta_{\sigma-} + v_\rho \cos 2\theta_{\rho-} + v_{\sigma} \cos 2 \theta_{\sigma-}.
\label{Vv}
\end{equation}
$v_e$ describes the backscattering of electrons $R\uparrow \rightarrow L\uparrow$ and $R\downarrow \rightarrow L\downarrow$.   $v_\rho$ describes the backscattering of a pair ($R\uparrow,R\downarrow \rightarrow L\uparrow,L \downarrow$), which also involves tunneling {\it spin} between the top and bottom edges.  $v_\sigma$ describes backscattering of spin
($R\uparrow,L\downarrow \rightarrow L\uparrow,R\downarrow$)
, which also involves tunneling {\it charge} between the top and bottom edges.   

The dimensions of these operators are 
\begin{equation}
\Delta(v_e) = (K + K^{-1})/2; \enskip \Delta(v_\rho) = 2 K; \enskip \Delta(v_\sigma) = 2 K^{-1}.
\label{dimensionv}
\end{equation}

Comparing (\ref{dimensiont},\ref{dimensionv}), it can be seen that they are the same with $\rho$ and $\sigma$ interchanged.  Indeed, the CC and the $II$ fixed points are precisely related by a reflection ${\cal M}$ that interchanges lead $2$ with lead $4$.  This exchanges the boundary conditions (\ref{iibc}) and (\ref{ccbc}) and takes $\varphi_\rho \leftrightarrow \theta_\sigma$ and $\varphi_\sigma \leftrightarrow \theta_\rho$.    Thus, the ${\cal M}$ operation takes a Hamiltonian expanded about $II$ to a Hamiltonian expanded about CC, with the identification:
\begin{align}
{\cal M}: \quad &  (\varphi_{\rho-},\varphi_{\sigma-}) \leftrightarrow (\theta_{\sigma-},\theta_{\rho-}); \\
& (t_e,t_\rho,t_\sigma) \leftrightarrow (v_e,v_\sigma,v_\rho); \\
&(K_\rho,K_\sigma) \leftrightarrow (K_\sigma^{-1},K_\rho^{-1}); \quad ( K \leftrightarrow K ).
\label{mtransform}
\end{align}

The ${\cal M}$ operation relates two different physical systems.   However, if a system is invariant under ${\cal M}$, then the two systems are identical.
${\cal M}$ is not a physical symmetry for a helical point contact because it exchanges the trivial and quantum spin Hall insulators.   In principle ${\cal M}$ could be a symmetry for helical modes on the surface of a topological crystalline insulator.   In either event, it is useful to consider the possibility of symmetry under ${\cal M}$ because that exchanges the II and CC phases.   Thus, for $1/2<K<2$, provided we assume that the II and CC phases are separated by a single critical point, then the presence of ${\cal M}$ symmetry tunes the system precisely to the unstable intermediate fixed point of interest.

When $K<1/2$ or $K>2$, (\ref{dimensiont},\ref{dimensionv}) imply that both the II and the CC fixed points are unstable.  These lead to strong coupling phases that will be identified as IC and CI, which we will see are also invariant under ${\cal M}$.   These phases will have a natural description in a dual theory described in the following section.

In addition to the mirror operation, we can also consider a change of variables that exchanges the charge and spin variables:
\begin{align}
{\cal E}: \quad&(\varphi_{\rho-},\varphi_{\sigma-}) \leftrightarrow (\varphi_{\sigma-},\varphi_{\rho-}) \\
&(t_e, t_\rho, t_\sigma)  \leftrightarrow (t_e, t_\sigma, t_\rho)  \\
&(K_\rho, K_\sigma) \leftrightarrow (K_\sigma,K_\rho); \quad (K \leftrightarrow K^{-1}).
\label{etransform}
\end{align}
This transformation relates the partition functions of two different problems with different values of $K$.  

\subsection{Constraints on the critical conductance}

The presence of mirror symmetry constrains the form of the conductance at the mirror invariant critical point.   Consider the conductance computed in an expansion about the II (CC) fixed point, $G_{IJ}^\alpha(K,{\bf t})$ ($G_{IJ}^\alpha(K,{\bf v})$), where ${\bf t} = (t_e,t_\rho,t_\sigma, ...)$ and ${\bf v} = (v_e,v_\rho,v_\sigma, ...)$.  Applying the combination of ${\cal M}$ and ${\cal E}$ it follows that 
\begin{equation}
G_{XX}^\rho(K,{\bf t}) = G_{YY}^\sigma(K^{-1},{\bf v} = {\bf t}).
\label{me}
\end{equation}

An additional relation follows from the fact that correlators of $\partial_x\theta_{\rho-}$ and correlators of $\partial_x\varphi_{\rho-}$ are related.   From (\ref{irhox},\ref{isigmay}), this implies that $G_{XX}^\rho(K)$ and $G_{YY}^\sigma(K)$ are related.   In Appendix \ref{appendixa} we show that the precise relation is
\begin{equation}
K^{-1} G^\rho_{XX}(K,{\bf t}) + K G^\sigma_{YY}(K,{\bf t}) = 2 e^2/h.
\label{gxxgyy}
\end{equation}

If we now assume that in the presence of symmetry under ${\cal M}$ the system flows to a unique mirror invariant fixed point with conductance $G_{IJ}^\alpha(K)^*$, then it follows by combining (\ref{me}) and (\ref{gxxgyy}) that
\begin{equation}
K^{-1} G_{XX}^\rho(K)^* + K G_{XX}^\rho(K^{-1})^* = 2e^2/h. \\
\label{gconstraint}
\end{equation}
Thus, the critical conductance obeys a symmetry under $K \leftrightarrow K^{-1}$.
A similar analysis shows that $G_{YY}^\rho$ obeys the same relation, and that $G^\sigma_{XX,YY}$ obey similar relations with $K$ replaced by $K^{-1}$.   

In Ref. \onlinecite{teokane09} it was also argued that for specific values of $K$ there is an additional symmetry that further constrains the conductance\cite{yikane98}.   The action $S_0^{II}[\varphi_{\rho-},\varphi_{\sigma-}]$ in (\ref{s0ii}) is invariant under rotations of $(\varphi_{\rho-},\varphi_{\sigma-}/K)$.   The cosine potential $V_{II}(\varphi_{\rho-},\varphi_{\sigma-})$ in ({\ref{Vt}) breaks this rotation symmetry, but when $t_e = 2t_\sigma$ and $t_\rho = 0$ symmetry under $C_6$ rotations is preserved. The dual action expanded about $S_0^{CC}$ has a similar symmetry.   Since the $C_6$ symmetry is preserved under the renormalization group, it is natural to expect that the intermediate critical point has the $C_6$ symmetry for $K=1/\sqrt{3}$.   A similar symmetry arises for $K= \sqrt{3}$, when $t_e = 2t_\rho$, $t_\sigma = 0$.   This enhanced symmetry predicts that for these values of $K$ the critical conductance satisfies
\begin{equation}
G_{XX}^\rho(K)^* = K^2 G_{XX}^\sigma(K)^*.
\end{equation}
Combined with (\ref{gconstraint}), this implies
\begin{equation}
G_{XX}^\rho(K)^* =  \left\{\begin{array}{ll}
(1/\sqrt{3})e^2/h & K = 1/\sqrt{3} \\
\sqrt{3} e^2/h & K = \sqrt{3}.
\end{array}\right.
 \label{gsqrt3}\end{equation}

\subsection{Web of Dualities}
\label{webofdualities}

Duality relations provide a connection between different representations of the same problem.   Problems with strong interactions that are therefore intractable in one representation can be weakly interacting, and therefore tractable, in a dual representation.   Duality relations have been a powerful tool for understanding the global behavior of quantum impurity problems, such as the problem of a single impurity in a Luttinger liquid.   In this section we identify several dual representations of the TK model.   This will be useful for identifying the mapping to the AG model, because the AG model, when expressed in terms of natural variables, is dual to the TK model.   The AG model exhibits the same dualities as the TK model, but in a different order.

We begin with the duality relating the CC and II limits\cite{kanefisher92b,yikane98}.   Starting from the II description in (\ref{hrho0},\ref{hsigma0},\ref{iibc},\ref{Vt}), when $t_e$ is large, the variables $\varphi_{\rho-}$ and $\varphi_{\sigma-}$ will be pinned in the deep minima of $-t_e \cos\varphi_{\rho-}\cos\varphi_{\sigma-}$ at $(\varphi_{\rho-},\varphi_{\sigma-})=(n_\rho \pi,n_\sigma\pi)$ with $n_\rho+n_\sigma$ even.  This implements the boundary conditions (\ref{ccbc}) up to a constant shift.  Fluctuations about these minima will involve instanton processes where $\varphi_{\rho-}$ and $\varphi_{\sigma-}$ jump between nearby minima.   Expanding the partition function in powers of these instantons is identical to expanding the partition function of (\ref{hrho0},\ref{hsigma0},\ref{ccbc},\ref{Vv}) in powers of $v_e$.   $v_e e^{i (\pm \theta_{\rho-}\pm\theta_{\sigma-})}$ generates an instanton where $(\varphi_{\rho-},\varphi_{\sigma-})$ jumps by  $(\pm \pi,\pm \pi)$.   

Thus, the same problem can be analyzed in two dual representations:  the $(\theta_{\rho-}, \theta_{\sigma-})$ representation in which the partition function is expanded in powers of the coefficient of the cosine potentials, $v$ or the $(\varphi_{\rho-},\varphi_{\sigma-})$ in which the partition function is expanded in powers of the fugacity of the instantons $t$.   Alternatively, one can view $t$ as the cosine potential, where $v$ the fugacity of instantons in $\varphi$.   We denote this duality transformation by
\begin{align}
{\cal D}_{\rho\sigma} : \quad  &(\varphi_{\rho-},\varphi_{\sigma-}) \leftrightarrow (\theta_{\rho-},\theta_{\sigma-}) \\
&(t_e, t_\rho, t_\sigma)  \leftrightarrow (v_e, v_\rho, v_\sigma)^{\rm dual} \\
&(K_\rho, K_\sigma) \leftrightarrow (K_\rho^{-1},K_\sigma^{-1}); \quad (K \leftrightarrow K^{-1}).
\end{align}
Here, the superscript ``dual" refers to the fact that $v$ and $t$ are inversely related:  large $v$ corresponds to small $t$ and vice versa.

\begin{figure}
\includegraphics[width=3in]{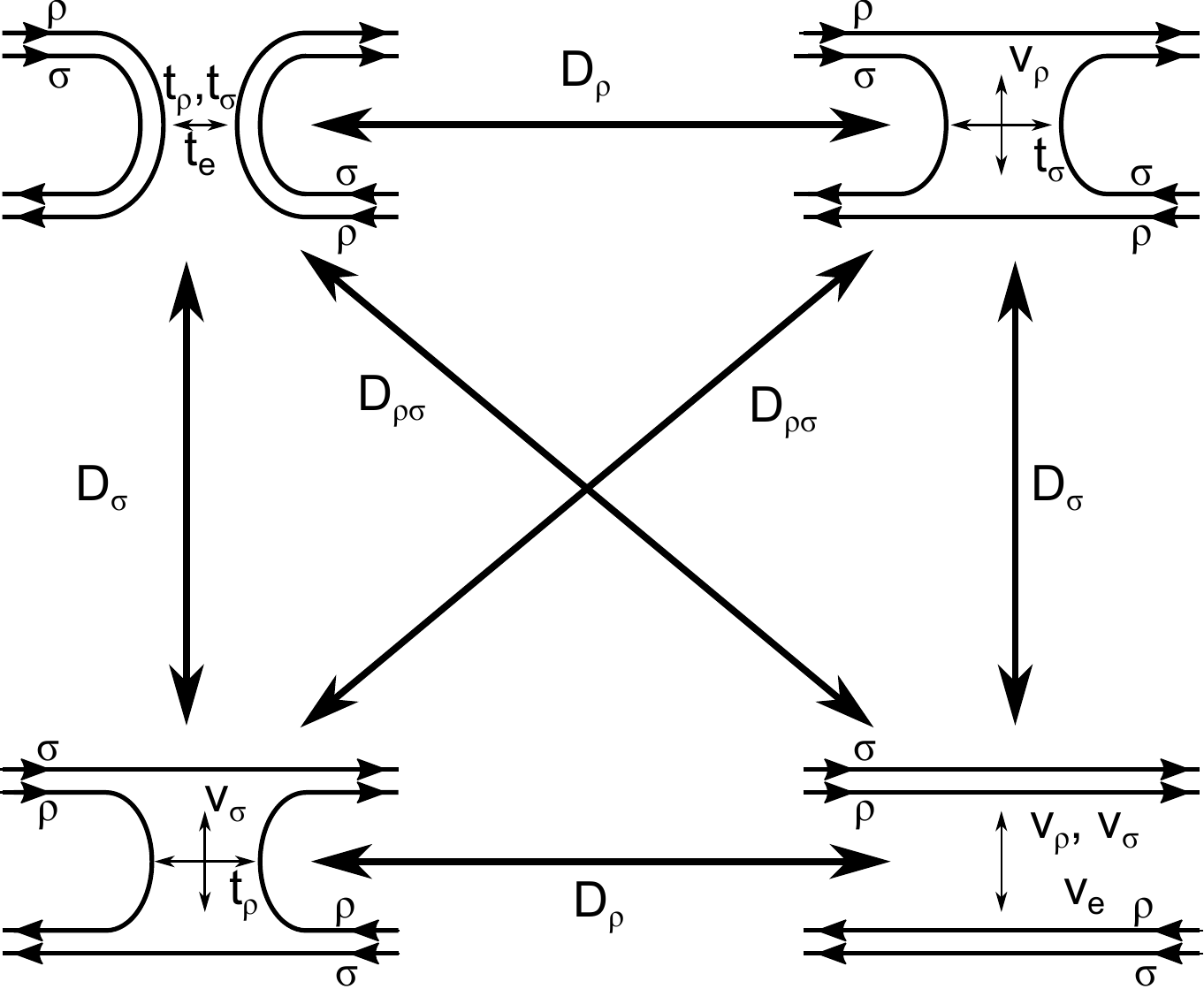}
\caption{Dualities relate four different representations of the helical point contact. }
\label{Fig2}
\end{figure}

In addition to the above duality transformation, one can also perform partial duality transformations independently in the charge or spin sectors.   These lead to a natural description of the charge insulator/spin conductor (IC) and charge conductor/spin insulator (CI) phases.    

Consider the II description in (\ref{hrho0},\ref{hsigma0},\ref{iibc},\ref{Vt}), and suppose that $t_\rho$ is large, while $t_e = t_\sigma = 0$.   Then $\varphi_{\rho-}$ will be pinned in the minima of $\cos 2\varphi_{\rho-}$ at $\varphi_{\rho-} = n_\rho \pi$, but $\varphi_{\sigma-}$ is free.   Up to a constant shift, this implements the mixed boundary condition
\begin{equation}
\varphi_{\rho-}(0) = 0; \quad \theta_{\sigma-}(0) = 0.
\end{equation}
There are two kinds of perturbations about this limit.   The first comes from $t_e \cos\varphi_{\rho-}\cos\varphi_{\sigma-}$, in which $\varphi_{\rho-}=n_\rho \pi$ is pinned.   The sign of this term, however, depends on the parity of $n_\rho$.   We therefore define the $Z_2$ variable $\tau_z = (-1)^{n_\rho}$.    The second kind of perturbation involves a tunneling process, in which $\varphi_{\rho-}$ jumps by $\pi$.   This process changes $n_\rho$ by 1 and flips the sign of $\tau_z$.    We can therefore represent these perturbations by the action $S^0_{IC} + \int d\tau V_{IC}(\tau)$ with
\begin{equation}
S_0^{IC} =  \int \frac{d\omega}{(2\pi)^2}\left[ K_\rho^{-1} |\omega||\theta_{\rho-}(\omega)|^2 + 
K_\sigma |\omega||\varphi_{\sigma-}(\omega)|^2\right]
\label{s0ic}
\end{equation}
and
\begin{equation}
V_{IC} = \tilde t_\sigma \tau_z \cos \varphi_{\sigma-}  + \tilde v_\rho \tau_x \cos \theta_{\rho-}
\label{vic}
\end{equation}
Note that
\begin{equation}
\Delta(\tilde t_\sigma) = K/2; \quad \Delta(\tilde v_\rho) = K/2.
\end{equation}
Both $\tilde t_\sigma$ and $\tilde v_\rho$ are irrelevant for $K>2$.  In that case, the IC fixed point is stable.    In addition, it is clear that the IC fixed point is invariant under the mirror operation ${\cal M}$.

A similar analysis can be applied to the CI fixed point.  For large $t_\sigma$, with $t_e=t_\rho=0$ we implement the boundary condition
\begin{equation}
\theta_{\rho-}(0)=0; \quad \varphi_{\sigma-}(0)=0.
\end{equation}
The expansion about this CI fixed point is generated by $S + S^0_{CI} + \int d\tau V_{CI}$ with
\begin{equation}
S_0^{CI} =  \int \frac{d\omega}{(2\pi)^2}\left[ K_\rho |\omega||\varphi_{\rho-}(\omega)|^2 + 
K_\sigma^{-1} |\omega||\theta_{\sigma-}(\omega)|^2\right]
\label{s0ci}
\end{equation}
and
\begin{equation}
V_{CI} = \tilde t_\rho \tau_z \cos \varphi_{\rho-}  + \tilde v_\sigma \tau_x \cos \theta_{\rho-}
\end{equation}
For $K<1/2$, $\tilde t_\rho$ and $\tilde v_\sigma$ are irrelevant.  The CI fixed point is stable and invariant under ${\cal M}$.

We thus have implemented the partial dualities:
\begin{align}
{\cal D}_\rho: \quad  & \varphi_{\rho-} \leftrightarrow \theta_{\rho-} \\
& (t_e,t_\sigma) \leftrightarrow (\tilde v_\rho^{\rm dual},\tilde t_\sigma) \\
& (K_\rho,K_\sigma)=(K,K^{-1}) \leftrightarrow (K_\rho^{-1},K_\sigma) = (K^{-1},K^{-1}).
\end{align}
and
\begin{align}
{\cal D}_\sigma: \quad  & \varphi_{\sigma-} \leftrightarrow \theta_{\sigma-} \\
& (t_e,t_\rho) \leftrightarrow (\tilde v_\sigma^{\rm dual},\tilde t_\rho) \\
& (K_\rho,K_\sigma)=(K,K^{-1}) \leftrightarrow (K_\rho,K_\sigma^{-1}) = (K,K).
\end{align}
Clearly, ${\cal D}_{\rho\sigma} = {\cal D}_\rho {\cal D}_\sigma$.   Thus, a given physical system can be described using four dual representations, as indicated in Fig. \ref{Fig2}.
In addition, the partial duality transformations interchange the ${\cal E}$ and ${\cal M}$ operations
\begin{equation}
{\cal E} = {\cal D}_\rho {\cal M} {\cal D}_\rho = {\cal D}_\sigma {\cal M} {\cal D}_\sigma.
\end{equation}

\section{Luttinger Liquid-Topological Superconductor Junction}
\label{section3}

\subsection{Model}

Affleck and Giuliano introduced a model of two semi-infinite single channel spinless Luttinger liquids coupled at their ends $x=0$ by single electron tunneling to the Majorana mode $\gamma$ of a 1D topological superconductor (whose phase we assume is pinned at $0$).  This leads to a tunneling Hamiltonian,
\begin{equation}
H_t = i t_1 \gamma (\Psi_1(0) + \Psi_1^\dagger(0)) + i t_2\gamma(\Psi_2(0) + \Psi_2^\dagger(0)).
\label{h0tag}
\end{equation} 
It is worth stressing that $H_t$ in 
(\ref{h0tag}) is symmetric under particle-hole exchange in only one channel. 
This forbids  any current flow between channels. However, such a symmetry is rather artificial; 
it could be broken by additional boundary interaction terms as,  for example, a  tunnelling term between channels. 

\begin{figure}
\includegraphics[width=3in]{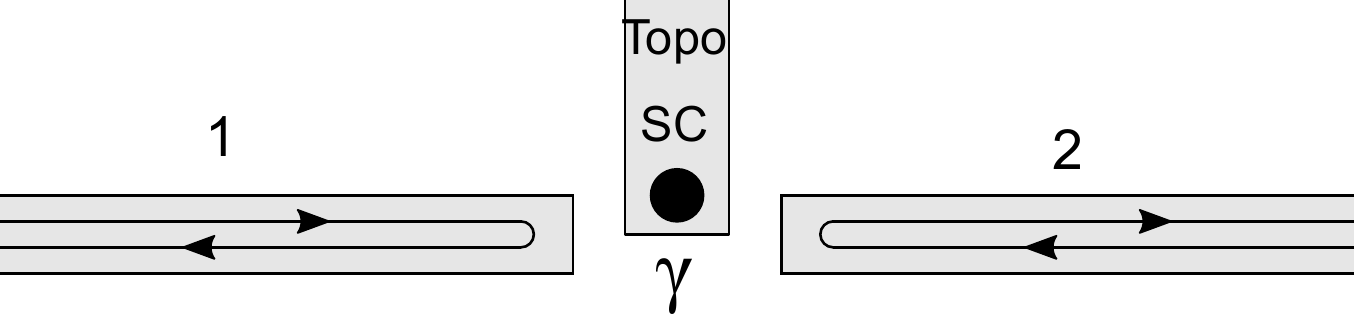}
\caption{Affleck Giuliano model of two Luttinger liquids coupled to the Majorana mode at the end of a 1D topological superconductor. }
\label{Fig3}
\end{figure}

Adopting the same notation as ({\ref{h0a}), we describe each Luttinger liquid by bosonizing, and writing
\begin{equation}
H = \int_0^\infty dx ({\cal H}_1^0 + {\cal H}_2^0)
\label{lut_AG1}
\end{equation}
with
\begin{equation}
{\cal H}_A^0 = \frac{v}{2\pi}\left[K_A^{-1} (\partial_x\theta_A)^2 + K_A(\partial_x\varphi_A)^2\right].
\label{lut_AG2}
\end{equation}
where $K_A$ is the Luttinger parameter characterizing the spinless Luttinger liquid in wires $A=1,2$.   We will assume that they are the same, 
\begin{equation}
K_1 = K_2 = K.
\end{equation}

The electron operator takes the form
\begin{equation}
\Psi_{A,p} = \Gamma_A \frac{e^{i(\varphi_A + p \theta_A)}}{\sqrt{2\pi x_c}}
\end{equation}
where $p = {\rm out,in} = +1,-1$.   
Here we have introduced Klein factors, represented by Majorana operators $\Gamma_1$ and $\Gamma_2$, which satisfy $\{\Gamma_A,\Gamma_B\} = 2\delta_{AB}$ and $\{\Gamma_A,\gamma\}=0$.

Defining the currents flowing into each contact as
\begin{equation}
I_A = \partial_t \theta_A/\pi = v K \partial_x\varphi_A/\pi
\end{equation}
we can define the Kubo conductances $G_{AB} = \lim_{\omega\rightarrow 0} \Pi_{AB}(\omega)$ with
\begin{equation}
\Pi_{AB}(t)=\theta(t)\langle[I_A(t),I_B(0)]\rangle.
\end{equation}
To facilitate comparison with the TK model, we also define a dual conductance $\tilde G_{AB}$ in terms of 
\begin{equation}
\tilde I_A = \partial_t\varphi_A/\pi = v K^{-1} \partial_x\theta_A/\pi.
\end{equation}
$\tilde G_{AA}$ describes the voltage response an applied current in lead A, and is related to $G_{AA}$.   Using the analysis in Appendix \ref{appendixa}, it follows that
\begin{equation}
K \tilde G_{AA} + K^{-1} G_{AA} = 2 e^2/h.
\end{equation}

\subsection{Phases}

In the absense of coupling to the superconductor, both wires have a boundary condition at the end $x=0$ that corresponds to perfect normal reflection for both wires (NN),
\begin{equation}
\theta_1(0)=0; \quad \theta_2(0) = 0.
\end{equation}
Integrating out the degrees of freedom for $x>0$, we can then consider the 0+1D action
for $\varphi_A(\tau) \equiv \varphi_A(x=0,\tau)$, given by
\begin{equation}
S_0^{NN} = \int \frac{d\omega}{(2\pi)^2} K |\omega| \left( |\varphi_1(\omega)|^2 + |\varphi_2(\omega)|^2 \right).
\label{s0nn}
\end{equation}
The single electron tunneling term (\ref{h0tag}) then has the form
\begin{equation}
V_{NN} = i t_1 \gamma \Gamma_1 \cos \varphi_1 + i t_2 \gamma \Gamma_2 \cos\varphi_2
\label{vnn}
\end{equation}
The scaling dimensions are
\begin{equation}
\Delta(t_1) = \Delta(t_2) = 1/(2K).
\end{equation}
In general, higher order, but less relevant, terms will also be present.   For example tunneling between the wires, $\cos (\varphi_1-\varphi_2)$ ($\Delta = 1/K$) and Josephson tunneling to the superconductor, $\cos 2\phi_a$  ($\Delta = 2/K$) are allowed.    The NN phase is stable when $K < 1/2$, while for $K>1/2$ $t_1$ and $t_2$ flow to strong coupling.

When $t_1$ is large, but $t_2$ is small, then $\varphi_1$ will lock into the minima of the cosine at $\varphi_1 = n\pi$ and $i\gamma\Gamma_1 = (-1)^n$.   In this limit, the Majorana mode is effectively absorbed by lead 1, changing the boundary condition at the end from perfect normal reflection to perfect Andreev reflection.   This is then described by the (AN) boundary conditions,
\begin{equation}
\varphi_1(0) = 0 ;\quad \theta_2(0) = 0
\end{equation}

This limit is described by a dual action of the form
\begin{equation}
S_0^{AN} = \int \frac{d\omega}{(2\pi)^2} |\omega| \left( K^{-1}|\theta_1(\omega)|^2 + K |\varphi_2(\omega)|^2 \right).
\label{s0an}
\end{equation} 
The electron operator into lead 1 now has the form $e^{\pm i \theta_1(0)}$, 
Perturbations about this limit include 
\begin{equation}
V_{AN} = t_{2J} \cos (2 \varphi_2) + v_e \cos \theta_1 \cos \varphi_2 + v_{1N} \cos 2 \theta_1
\label{van}
\end{equation}
$t_{2J}$ can be interpreted as tunneling a Cooper pair into the superconductor from lead 2.   $v_e$ describes two processes, in which an electron is tunneled from one lead to the other, or where a Cooper pair is removed from the superconductor, adding one electron to each lead.   $v_{1N}$ describes the normal reflection of an electron in lead 1.   These operators have dimensions
\begin{equation}
\Delta(t_{2J}) = 2/K; \quad \Delta(v_e) = (K + K^{-1})/2; \quad \Delta(v_{1N}) = 2K
\end{equation}
An identical analysis can be applied to the NA fixed point by interchanging leads 1 and 2.  

It is clear that both the NA and the AN fixed points are stable for $1/2 < K < 2$.  Affleck and Giuliano observed that when $t_1=t_2$, the system must flow to the critical point that separates them, which is described by an unstable intermediate fixed point.   When $t_1=t_2$, the AG model has a reflection symmetry ${\cal M}$, which interchanges leads 1 and 2,
\begin{align}
{\cal M}: \quad &  (\varphi_{1},\theta_{1}) \leftrightarrow (\varphi_{2},\theta_{2}); \\
& (t_1,t_2) \leftrightarrow (t_2,t_1); \\
&(K_1,K_2) \leftrightarrow (K_2,K_1); \quad ( K \leftrightarrow K ).
\end{align}
Note the similarity and difference with (\ref{mtransform}).  In contrast to the TK model, ${\cal M}$ symmetry in the AG model is implemented by a physical reflection symmetry of the junction.

For $K<1/2$, $v_{1N}$ is relevant at the AN fixed point and flows to strong coupling.   Likewise, $v_{2N}$ is relevant at the NA fixed point.   This suggests that in either case the system flows to the stable NN fixed point.    For $K>2$, $t_{2J}$ ($t_{1J}$) is relevant at the AN (NA) fixed point.   This suggests that lead 2 becomes strongly coupled to the superconductor, and the system flows to a AA phase where both leads are Andreev reflected from the superconductor.   

We therefore see that the stable phases of the AG model have a structure very similar to that of the TK model.  In the following section we will describe the precise translation between these two models.

\section{Equivalent Models}
\label{section4}

We now establish the equivalence between the TK model and the AG model.  If we compare Eqs. \ref{s0ic} and \ref{vic} for the TK model expanded about the IC limit with Eqs.  \ref{s0nn} and \ref{vnn} of the AG model expanded about the NN limit it can be seen that they are identical, provided the products of Majorana operators in (\ref{vnn}) are represented by Pauli matrices: 
\begin{equation}
i \gamma \Gamma_1 \leftrightarrow \tau_z, \quad
i \gamma \Gamma_2 \leftrightarrow \tau_x,
\end{equation}
and the boson variables are related by 
\begin{align}
(\theta_1,\varphi_1)&\leftrightarrow (\theta_{\rho-},\varphi_{\rho-}),\\
(\theta_2,\varphi_2)&\leftrightarrow (\varphi_{\rho-},\theta_{\rho-}).
\end{align}
Expanding the actions in powers of the cosine perturbations and integrating out the boson fields  generates identical 0+1D Coulomb gas expansions in for the two problems, provided the Luttinger parameters are related by
\begin{equation}
K_1 \leftrightarrow K_\rho, \quad
K_2 \leftrightarrow 1/K_\sigma.
\end{equation}
With this correspondence, we can relate the Kubo conductances
\begin{align}
&G_{11} \leftrightarrow G^\rho_{XX}, \quad
G_{22}\leftrightarrow G^\rho_{YY} , \\
&\tilde G_{22} \leftrightarrow G^\sigma_{XX} , \quad
\tilde G_{11} \leftrightarrow G^\sigma_{YY}.
\end{align}

This precise equivalence relating the NN limit of the AG model to the IC limit of the TK model suggests that there is a correspondence between all of the phases and fixed points of both models,
\begin{align}
&{\rm NN}\leftrightarrow {\rm IC}, \quad {\rm AN}\leftrightarrow {\rm CC}, \\
&{\rm NA}\leftrightarrow {\rm II},\ \quad {\rm AA}\leftrightarrow {\rm CI}.
\end{align}
In the following, we explore in detail different aspects of the correspondence, including the correspondence of dualities, 
and the correspondence between the free fermion limits of both models, which provides insight into the mathematical structure underlying the equivalence between these rather different models.

\subsection{Dualities}

The defining action for the TK model, given by $S_{CC}$ (or $S_{II}$), is expressed in terms of natural variables $(\theta_{\rho-},\theta_{\sigma-})$ (or $(\varphi_{\rho-},\varphi_{\sigma-})$), while the defining action for the AG model, $S_{NN}$,  is expressed in terms of $(\varphi_1,\varphi_2)$.   This indicates that the defining representations of the two models are related by the ${\cal D}_\sigma$ (or ${\cal D}_\rho$) partial duality.   

Given this identification, both models exhibit the same web of dualities discussed in Section \ref{webofdualities}.  Indeed, it is easy to see that the AN and NA limits of the AG model are related by the ${\cal D}_{\rho\sigma}$ type duality by, for instance, considering the limit of large $(t_J, v_e, v_{1N})$ in which pins $\theta_1$ and $\varphi_2$ (\ref{van}), and expanding in powers of instantons.   In addition, an instanton analysis similar to the ${\cal D}_\rho$ and ${\cal D}_\sigma$ duality relates the NA limit to the NN and AA limits.

A case that is somewhat less obvious is the duality between the NN and the AA limits of the AG model (or equivalently between IC and CI of the TK model).  Due to the Majorana operators, the two terms in (\ref{vnn}) do not commute with each other.   Therefore, one can not straightforwardly perform the instanton analysis in limit of large $t_1$ and $t_2$.   
Here we present two simple modifications of this argument allows the analysis to proceed.   

Consider first an extension of the AG model, in which  Josephson tunneling of Cooper pairs between the superconductor and the leads, described by
\begin{equation}
V_J = t_{1J} \cos 2\varphi_1 + t_{2J} \cos 2\varphi_2
\end{equation}
is included in (\ref{vnn}).   Consider first the case in which the single particle tunneling vanishes, $t_1=t_2=0$.    $K>2$, both $t_{1J}$ and $t_{2J}$ are relevant and flow to strong coupling, leading to a phase in which $\varphi_1$ and $\varphi_2$ are pinned.   This is described by a dual action
\begin{equation}
S_0^{AA} = \int \frac{d\omega}{(2\pi)^2} K^{-1} |\omega| \left( |\theta_1(\omega)|^2 + |\theta_2(\omega)|^2 \right).
\label{s0aa}
\end{equation}
Now we can add single electron coupling to the Majorana mode of the superconductor, (\ref{h0tag}).  In this case, since $\varphi_a$ is pinned the tunneling term takes the form,
\begin{equation}
V_{AA} = i v_1 \gamma \Gamma_1 \cos \theta_1 + i v_2 \gamma \Gamma_2 \cos\theta_2.
\label{vaa}
\end{equation}
These perturbations have dimension
\begin{equation}
\Delta(v_1) = \Delta(v_2) = K/2,
\end{equation}
and are irrelevant for $K>2$.   

A slight modification of this argument is to consider single-electron tunneling to a topological superconductor that has {\it three} low energy Majorana modes $\gamma_0$, $\gamma_1$, $\gamma_2$, described by a tunneling Hamiltonian
\begin{equation}
V_{NN} = \sum_{a = 1}^2 \sum_{j = 0}^2  i t_{aj} \Gamma_a \gamma_j  \cos\varphi_a.
\end{equation}
While the existence of three low energy Majorana modes is not generic, it is always possible to have extra low energy Andreev bound states, and the two extra ones do not need to be exactly at zero energy.    Consider the limit of large $t_{11}$ and $t_{22}$, which fix $i\Gamma_a \gamma_a = \pm 1$, and $\varphi_a = n_a \pi$ for $a=1,2$.   Now, instantons in which $\varphi_a$ jumps by $\pi$ and $i\Gamma_a \gamma_a$ changes sign are generated by precisely (\ref{s0aa},\ref{vaa}), with $\gamma \rightarrow \gamma_0$.

  An alternative pathway towards the derivation of (\ref{vaa}),
which we review in Appendix \ref{k2e}, is based on a lattice fermion description of the system, which eventually 
yields  (\ref{s0aa},\ref{vaa})  in the continuum limit \cite{giuliano19}. 

\subsection{Free fermion limit}\label{freefermionsec}

Non interacting electrons are described by the $K=1$ limit in both the TK model and the AG model.  In both cases, the low energy behavior is described by a fixed line with a continuously varying transmission matrix.   While the $K=1$ limits of both models coincide, the mapping between them is nontrivial.    In this section we outline the precise connection between the free fermion limits of both models.

We begin by considering the free fermion limit of the TK model.   In this case there are four incoming and four outgoing Dirac fermion modes, related by a unitary scattering matrix.  Single electron scattering states are combinations of incoming and outgoing waves related by
\begin{equation}
\psi_{a,\rm out} = \sum_b {\cal S}^{TK}_{ab} \psi_{b,\rm in}
\end{equation}
In the spin and charge conserving model that we consider here (where incoming leads 1 and 3 and outgoing leads 2 and 4 have spin up), the S-matrix has the form\cite{teokane09} 
\begin{equation}
{\cal S}^{TK} = \left(\begin{array}{cccc} 
0 & t_1 & 0 & r_1 \\
t_2 & 0 & r_2^* & 0 \\
0 & r_1^* & 0 & -t_1^* \\
r_2 & 0 & -t_2^* & 0
\end{array}\right),
\end{equation}
with $|r_1|^2 + |t_1|^2 = |r_2|^2 + |t_2|^2=1$.   
This can be interpreted as the transmission and reflection of up and down spins, described by a $SU(2)\oplus SU(2)$ scattering matrix, with 6 independent real degrees of freedom.    In addition, one could introduce two additional $U(1)$ phases for the up and down spins by multiplying $S^{TK}$ by ${\rm diag}(e^{i\zeta_1},e^{i\zeta_2},e^{i\zeta_1},e^{i\zeta_2})$.   We will see that these phases only affect the total charge ($\rho+$) and total spin ($\sigma+$) sectors, and cancel in our analysis.  
Time reversal symmetry in the TK model imposes a further constraint $S = - Q S^T Q$ where $Q = {\rm diag}(1,-1,1,-1)$\cite{teokane09}, which in this case requires $t_1=t_2$ and $r_1 = r_2$.   Here, we will consider the somewhat more general case where time reversal symmetry can be violated, but $S_z$ conservation is preserved.
While this is not a natural symmetry for the TK model, it is useful to consider time reversal breaking in the AG model.

For $K=1$ the AG model is likewise expressed in terms of free fermions.   In this case there are two incoming and two outgoing free fermion channels, which can be related by either normal or Andreev scattering.   This is described by a unitary $4\times 4$ scattering matrix for the Nambu spinor $(\Psi_1, \Psi_2, \Psi_1^\dagger, \Psi_2^\dagger)^T$, 
\begin{equation}
{\cal S}^{AG} = \left(\begin{array}{cccc}
n_{1,1} & n_{1,2} & a_{1,1} & a_{1,2} \\
n_{2,1} & n_{2,2} & a_{2,1} & a_{2,2} \\
a^*_{1,1} & a^*_{1,2} & n^*_{1,1} & n^*_{1,2} \\
a^*_{2,1} & a^*_{2,2} & n^*_{2,1} & n^*_{2,2}
\end{array}\right).
\label{sag}
\end{equation}
This obeys the Bogoliubov de Gennes constraint ${\cal S}^{AG*} = \tau_x {\cal S}^{AG} \tau_x$.   The amplitudes $n_{A,B}$ ($a_{A,B}$) for Normal (Andreev) transmission from channel $A$ to channel $B$ are not all independent because they are constrained by unitarity, ${\cal S}^{AG\dagger} {\cal S}^{AG} = 1$.   The counting of degrees of freedom is simplest if one expresses $\Psi_A$, $\Psi_A^\dagger$ in terms of four Majorana modes as $\xi_A \pm i \eta_A$.   In that basis, the scattering matrix is a real orthogonal $4\times 4$ matrix.   A further constraint follows from the fact that there are two distinct topological classes of orthogonal scattering matrices distinguished by the sign of their determinant.  It is well known that these two classes correspond to whether or not there is a  Majorana zero mode at the scattering center\cite{fulga11,fulga11b}.   Thus, for the AG model we expect
\begin{equation}
{\rm det}[{\cal S}^{AG}] = -1.
\end{equation}  
Therefore ${\cal S}^{AG}$ is an improper rotation times a $SO(4)$ matrix.  Since $SO(4) \sim SU(2) \oplus SU(2)$, with 6 real degrees of freedom, it is plausible that ${\cal S}^{AG}$ and ${\cal S}^{TK}$ are related.   In the following we will deduce the explicit relation.

Consider the bosonized representations of both models.   The TK model is built out of ``TK fermions", represented by
\begin{equation}
\psi_{a,p} \sim \gamma_a e^{i \phi_{a,p}}
\label{psitk}
\end{equation}
Here, as in (\ref{psibosonize}), we express the incoming/outgoing ($p=+1/-1$)  electron operator in channel $a$ in terms of a chiral boson $\phi_{a,p}$.   Here we also keep the Klein factor, which satisfies $\{\gamma_a,\gamma_b\}= 2\delta_{ab}$.   Note that the product $(-1)^F = \gamma_1\gamma_2\gamma_3\gamma_4 = \pm 1$ describes the fermion parity and commutes with the Hamiltonian.

The AG model is built from ``AG fermions",
\begin{align}
\Psi_{1,p} = &\Psi_{\rho-,p} \sim \Gamma_1 e^{i (\phi_{1,p} - \phi_{2,p} - \phi_{3,p} + \phi_{4,p})/2} \label{Psi1ag} \\
\Psi_{2,p} = &\Psi_{\sigma-,p}^p \sim \Gamma_2 e^{i (\phi_{1,p} + \phi_{2,p} - \phi_{3,p} - \phi_{4,p})/2}.\label{Psi2ag} 
\end{align}
In addition, we can define two additional fermions that are not present in the AG model,
\begin{align}
 \Psi_{3,p} = &\Psi_{\rho+,p} \sim \Gamma_3 e^{i(\phi_{1,p} + \phi_{2,p} + \phi_{3,p} + \phi_{4,p})/2},\label{Psi3ag} \\
 \Psi_{4,p} = &\Psi_{\sigma+,p}^p \sim \Gamma_4 e^{i(\phi_{1,p} - \phi_{2,p} + \phi_{3,p} - \phi_{4,p})/2}.\label{Psi4ag} 
\end{align}
A special property of this transformation is that it preserves the commutation relations obeyed by the chiral fields, which guarantees that both $\psi_a$ and $\Psi_A$ are fermion operators.   However, due to the 1/2 in (\ref{Psi1ag}-\ref{Psi4ag}) the AG fermions, $\psi_a$ are related to the TK fermions, $\Psi_A$, by a non-local transformation.  

The fact that this is possible is related to the {\it triality} of $SO(8)$.  The four free fermion channels of the TK model can be expressed in terms of 8 Majorana fermions $\psi_a = \xi_a + i \eta_a$.   In this basis, the transmission problem is expressed in terms of a $SO(8)$ scattering matrix that is subject to the constraints of charge and spin conservation.   In this representation, the Majorana operators $\xi_a$, $\eta_a$ transform under the fundamental 8-dimensional vector representation of $SO(8)$, and the scattering matrix ${\cal S}^{TK}$ is a representation of a $SO(8)$ rotation in that fundamental representation.

In addition to the 8-dimensional vector representation, $SO(8)$ has two 8-dimensional spinor representations.   This follows from the fact that a the Clifford algebra $Cl_{0,8}$ can be represented in terms of 8 $16\times 16$ real Dirac matrices.  There is also a 9th Dirac matrix $``\gamma^5" = \prod_{i=1}^8 \gamma_i$, whose eigenvalue $\pm 1$ distinguishes the two 8-dimensional Majorana-Weyl spinor representations.
The vector representation and the two spinor representations of $SO(8)$ are  related by triality symmetry.   $SO(8)$ triality has also appeared in other contexts, including the Gross Neveau Model\cite{shankar80}, the Ashkin Teller model\cite{shankar85} the two channel Kondo problem\cite{maldacena97} and the theory of the two leg Hubbard ladder\cite{balents98}.

While the TK fermions $\psi_a$ (expressed in terms of Majorana operators) transform in the fundamental vector representation, the AG fermions, $\Psi_{1,2}$ and $\Psi_{1,2}^\dagger$, together with the extra fields $\Psi_{3,4}$ and $\Psi_{3,4}^\dagger$, transform in one of the spinor representations.   The scattering matrix ${\cal S}^{AG}$ is a representation of the same $SO(8)$ rotation as ${\cal S}^{TK}$ in the spinor representation, similar to representing $SO(3)$ rotations in terms of $SU(2)$ spinors.

Due to charge and spin conservation, the components of the scattering matrix describing $\Psi_{3}$ and $\Psi_{4}$ are trivial and describe perfect transmission (with phase shifts determined by $\zeta_1 \pm \zeta_2$).
To extract the explicit form of ${\cal S}^{AG}$ in terms of ${\cal S}^{TK}$ it is simplest to consider the transmission of pairs of electrons.   Due to the $1/2$ in (\ref{Psi1ag}-\ref{Psi4ag}), a single AG fermion is related by a non-local transformation to the TK fermions.   However, an AG fermion combined with a $\Psi_3 = \Psi_{\rho+}$ fermion is related locally to a pair of TK fermions.   Explicitly, we have
\begin{align}
&\Psi_{3}\Psi_1 = \psi_4 \psi_1,  \quad
&\Psi_{3}\Psi_1^\dagger = \psi_2 \psi_3, \label{psitranslate1}\\
&\Psi_{3}\Psi_2 = \psi_1 \psi_2, \quad \label{psitranslate2}
&\Psi_{3}\Psi_2^\dagger = \psi_4 \psi_3. 
\end{align}
Note that since $\psi_a$ anticommute, there is some freedom in choosing the signs in (\ref{psitranslate1},\ref{psitranslate2}).   However, they are not all independent, and depend on the product of Klein factors $\gamma_1\gamma_2\gamma_3\gamma_4 = \pm 1$.   Eq. \ref{psitranslate1} implies $\gamma_4\gamma_1=\gamma_2\gamma_3$, which is consistent with (\ref{psitranslate2}), which implies $\gamma_1\gamma_2 = \gamma_4\gamma_3$.

Using, (\ref{psitranslate1},\ref{psitranslate2}) we can express two particle scattering states in terms of either TK fermions or AG fermions.   When expressed in terms of TK fermions, the incoming states $\psi_{a,{\rm in}}$, $\psi_{b,{\rm in}}$ and outgoing states $\psi_{c,{\rm out}}$, $\psi_{b,{\rm out}}$ will be expressed in terms of the product of single particle scattering matrices: ${\cal S}_{ac}^{TK} {\cal S}_{bd}^{TK}-{\cal S}_{ad}^{TK}{\cal S}_{bc}^{TK}$.   On the other hand, since $\Psi_3$ is transmitted perfectly, the same process is described by ${\cal S}^{AG}_{\alpha\beta}$, where $\alpha,\beta$ are related to $ab, cd$ as in (\ref{psitranslate1},\ref{psitranslate2}).  
This leads to
\begin{equation}
{\cal S}^{AG} = \left(\begin{array}{cccc}
-r_1 r_2 &   t_1 r_2  & t_1 t_2^*  & r_1 t_2^* \\
 r_1 t_2  &  -t_1 t_2  & t_1 r_2^*  & r_1 r_2^* \\
 t_1^* t_2 &  r_1^*t_2 &-r_1^*r_2^* &t_1^*r_2^* \\
t_1^*r_2 & r_1^*r_2 & r_1^*t_2^* &-t_1^*t_2^*
\end{array}\right).
\label{sss}
\end{equation}
This result is independent of the $U(1)$ phases $e^{i(\zeta_1+\zeta_2)}$, which are canceled by the phase shift for transmission of $\Psi_3$. 
This parameterization satisfies unitarity ${\cal S}^{AG\dagger}{\cal S}^{AG}=1$, ${\cal S}^{AG*} = \tau_x {\cal S}^{AG}\tau_x$ as well as ${\rm det}[S^{AG}]=-1$, and is the most general form satisfying those constraints.   

The transmission in the TK model can be characterized by the transmission and reflection probablities,
\begin{align}
T_a &= |t_a|^2 \label{t12}\\
R_a&= |r_a|^2 = 1-T_a
\end{align}
for $a=1,2$.
Likewise the transmission in the AG model can be characterized by the probabilities for normal and Andreev reflection,
\begin{align}
N_{i,j} &= |n_{i,j}|^2 \\
A_{i,j} &= |a_{i,j}|^2.
\end{align}
From (\ref{sss}) it can be seen that each of these is a product of two $R$'s or $T$'s:   
\begin{align}
N_{1,1}=A_{2,2} &= R_1 R_2 \label{c.1}\\
N_{1,2}=A_{2,1} &=  T_1 R_2 \label{c.2} \\
N_{2,1}=A_{1,2} &=  R_1 T_2 \label{c.3} \\
N_{2,2}=A_{1,1} &= T_1 T_2. \label{c.4}
\end{align}

It is instructive to compare the Landauer conductances.
In the time reversal invariant ($T_1 = T_2 \equiv T$) case we have
\begin{equation}
G^\rho_{XX} = 2 T e^2/h
\end{equation}
In the AG model, we have
\begin{equation}
G_{11} = (1 - N_{1,1} + A_{1,1}) e^2/h
\end{equation}
Using the fact that $A_{1,1} = T^2$ and $N_{1,1} = R^2 = (1-T)^2$, it follows that $G_{11} = G^\rho_{XX}$.  A similar analysis can also be applied to the other components of the conductance.

\subsection{Renormalization of ${\cal S}^{AG}$ for weak interactions}

As a non-trivial check of the equivalence between the TK model and the AG model we perform an analysis of the renormalization group flow of the scattering matrix $S^{AG}$ for weak interactions, $K=1-\epsilon$.   This type of analysis was introduced by Matveev, Yue and Glazman\cite{matveev93,yue94}, who studied the single impurity problem in a spinless Luttinger liquid.  They found that due to the interference between the incident and reflected waves the scattering matrix is renormalized to linear order in $\epsilon$ leading to a renormalization group flow towards perfect reflection for repulsive interactions, ($\epsilon>0$).   

TK performed a similar analysis for the helical point contact.   Since time reversal symmetry forbids reflection, it was found that the renormalization of the scattering matrix to linear order in $\epsilon$ vanishes.   However, there is a correction at order $\epsilon^2$, which allowed TK to compute the universal crossover from the critical point to the II and the CC fixed points.

Here we do the corresponding calculation for the AG model.   The mapping outline in the previous section suggests that there should be a correspondence between the renormalization group flows in the two models.   However, the structure of the problems is different, because in the AG model there is no constraint on the normal and Andreev reflection.   Therefore, in general, one should expect a renormalization to linear order in $\epsilon$.   We will show that for the class of scattering matrices with ${\rm det}[S^{AG}] = -1$, the linear in $\epsilon$ renormalization vanishes and that the $\epsilon^2$ renormalization is in agreement with TK.   Our analysis also generalizes the result of TK to include the case where time reversal symmetry is violated, but spin conservation is preserved.

In Appendix \ref{rg_appe} we derive the renormalization group flow for ${\cal S}^{AG}$.  Here we report the result, and drop the superscript $AG$ for brevity.   To linear order in $\epsilon = 1-K$, we find
\begin{equation}
\frac{d{\cal S}_{ab}}{d\ell} = \frac{\epsilon}{2} ( v_{ab} {\cal S}_{ab} - \sum_{c,d} v_{cd} {\cal S}_{ad}{\cal S}^*_{cd}{\cal S}_{cb} ).
\label{linearepsilon}
\end{equation}
Here
\begin{equation}
v_{ab} = \delta_{a,b} - \delta_{a,b+2}
\end{equation}
describes the interaction $\psi_{a,{\rm in}}^\dagger\psi_{a,{\rm in}} \psi_{a,{\rm out}}^\dagger \psi_{a,{\rm out}}$.   In the second term, which accounts for the BdG form of the Hamiltonian, $a$ and $b$ are understood to be defined modulo 4.

The TK calculation has a similar structure, except $v_{ab} = \delta_{a,b}$, and time reversal symmetry requires ${\cal S}_{aa} = 0$.   Thus, for ${\cal S}^{TK}$ the right hand side of (\ref{linearepsilon}) vanishes.   

For a general S-matrix of the form (\ref{sag}), the right hand side of (\ref{linearepsilon}) does not vanish.   However, as detailed in Appendix \ref{rg_appe} for the class of BdG S-matrices with ${\rm det}[{\cal S}] = -1$, the right hand side of (\ref{linearepsilon}) {\it does} vanish.   This can be checked by plugging the parameterization (\ref{sss}) into (\ref{linearepsilon}).

To second order in $\epsilon$ the renormalization of ${\cal S}$ is given by,
\begin{align}
\frac{d{\cal S}_{ab}}{d\ell} =\frac{\epsilon^2}{4} &(
\sum_{cd} v_{ad} v_{cb}  {\cal S}_{ab} {\cal S}_{cd}{\cal S}_{cd}^*  - \nonumber\\
&\sum_{cdef} v_{cf}v_{ed} {\cal S}_{ad}{\cal S}_{ef}{\cal S}^*_{cd}{\cal S}^*_{ef}{\cal S}_{cb}).
\label{epsilonsquare}
\end{align}
For $v_{ab} = \delta_{ab}$, this has a structure similar to the TK calculation.   If we plug the parameterization (\ref{sss}), then (\ref{epsilonsquare}) can be expressed in terms of $T_1$ and $T_2$, in (\ref{t12}),
\begin{align}
\frac{dT_1}{d\ell} = -\epsilon^2 T_1 (1 - T_1)(1 - 2 T_2) \\
\frac{dT_2}{d\ell} = -\epsilon^2 T_2 (1 - T_2)(1 - 2 T_1).
\label{rgflow}
\end{align}

If time reversal symmetry is present, so that $T_1=T_2=T$, then this result agrees precisely with the TK result for the case in which spin is conserved.   In this case, there are two stable points:  $T=0$ corresponds to the NA fixed point with $N_{1,1}=A_{2,2}=1$, while $T=1$ corresponds to the AN fixed point with $N_{2,2}=A_{1,1}=1$.   In addition, the unstable critical fixed point at $T=1/2$ corresponds to $N_{i,j}=A_{i,j} = 1/4$ for all $i,j=1,2$.

When time reversal symmetry is violated, the renormalization group flows for $T_1$and $T_2$ are shown in Fig. \ref{Fig4}.   Two additional unstable fixed points occur at $T_1=1$, $T_2=0$  ($N_{1,2} = A_{2,1} = 1$) and $T_1=0$, $T_2=1$ ($N_{2,1}=A_{1,2}=1$).   The critical point separating the two stable phases is on the time reversal symmetric line, as is the universal crossover from the critical fixed point to the stable fixed points.

\begin{figure}
\includegraphics[width=2in]{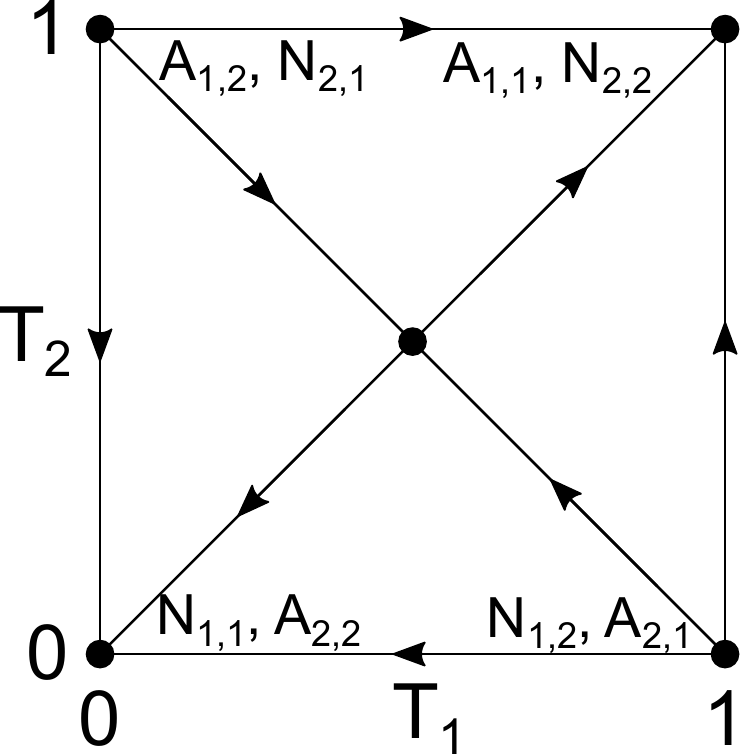}
\caption{Flow diagram for the order $\epsilon^2$ RG flow of ${\cal S}^{AG}$ parameterized using (\ref{sss}). }
\label{Fig4}
\end{figure}

\section{Predictions for the AG model}
\label{section5}

Having established the correspondence between the AG model and the TK model, we now collect some new results for the AG model, focusing on the behavior of the conductances $G_{11}$ and $G_{22}$ of the AG model.   The results are a direct translation of the predictions of TK, except that TK computed the ``physical conductance" that is modified to account for Fermi liquid leads.   Here, in order to facilitate comparison with numerics, we focus on the Kubo conductance, which is computed using the Kubo formula for infinite Luttinger liquid leads.
The translation between these is reviewed in Appendix \ref{appendixc}.

Consider first the conductance at the mirror symmetric critical point $G^*_{11}(K) = G^*_{22}(K)$.   For $K<1/2$ ($K>2$) the NN (AA) fixed point is stable, so that the Kubo conductance is $0$ ($2 K e^2/h)$.  For $1/2<K<2$, the critical conductance varies between $0$ and $2 K e^2/h$.   The results are simplest if we define
\begin{equation}
G^*_{11}(K) = G^*_{22}(K) = 2 K \mu^*(K) e^2/h .
\end{equation}
From Eq. (\ref{gconstraint}), the normalized Kubo conductance $\mu^*(K)$ satisfies
\begin{equation}
\mu^*(K) + \mu^*(K^{-1}) = 1.
\label{gstarsymmetry}
\end{equation}
TK computed the critical physical conductance perturbatively for $K = 1-\epsilon$, $K= 2-\epsilon$, $K=1/2+\epsilon$ as well as at $K = \sqrt{3}$ and $K=1/\sqrt{3}$.   Here we translate those results to the normalized Kubo conductance in the AG model.

For $K=1$, the conductance at the critical fixed point is $G^*=e^2/h$.  For weak interactions TK found that the correction to the physical conductance linear in $\epsilon$ is equal to zero.   Using Eq. \ref{kubotophysicalag}, this implies
\begin{equation}
\mu^*(1-\epsilon) = 1/2 + \epsilon/4.
\end{equation}
The vanishing correction occurred for TK because time reversal symmetry required the reflection amplitude vanish for $K=1$.   Including the interactions perturbatively, the only possible non-zero diagrams that contributed to the conductance were the ``random phase approximation (RPA)" type bubble diagrams.  While these contribute to the Kubo conductance, their contribution to the physical conductance was zero.   

The perturbative calculation of the conductance in the AG model is somewhat more complicated.   In addition to the RPA diagram, there are additional non-zero diagrams that contribute at first order in the interactions.   Nonetheless, we have checked that the non-RPA diagrams cancel, and the two calculations agree.

For $K$ close to $1/2$ the critical fixed point is close to the NA fixed point, allowing for a perturbative calculation of the conductance.   TK found that the physical conductance is $2\pi^2 \epsilon e^2/h$ for $K=1/2+\epsilon$, which translates to\footnote{This result was also found for the AG model in Ref. \onlinecite{affleck13}, but was changed in a subsequent revision\cite{affleck19}.  There was an error in the revision, and the original result stands, in agreement with Eq. \ref{gstar1} and Refs. \onlinecite{kanefisher92b,teokane09}.}
\begin{equation}
\mu^*(1/2+\epsilon) =  2\pi^2 \epsilon.
\label{gstar1}
\end{equation}
Similarly, for $K = 2-\epsilon$, using (\ref{gstarsymmetry}) we have
\begin{equation}
\mu^*(2-\epsilon) = 1 -  \pi^2 \epsilon/2.
\label{gstar2}
\end{equation}

\begin{figure}
\includegraphics[width=3in]{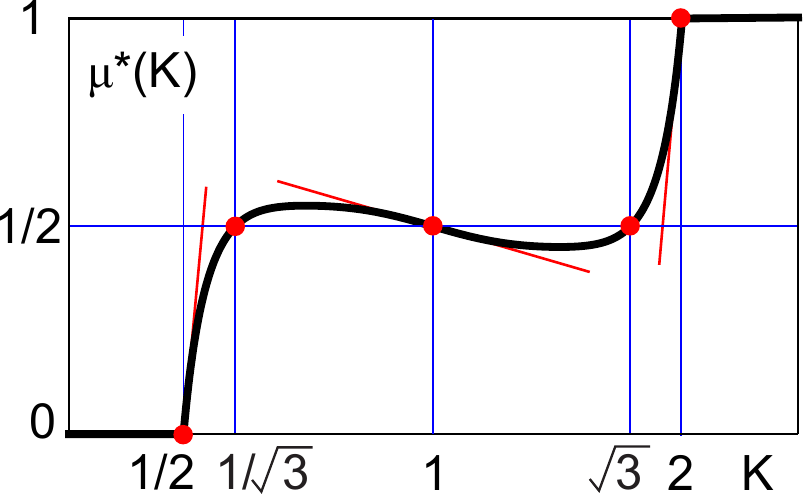}
\caption{Schematic plot of the critical conductance $G_{11} = G_{22} = G^*(K)$ in the AG model, expressed in terms of $\mu^*(K)=G^*(K)/(2Ke^2/h)$.   }
\label{Fig5}
\end{figure}

Finally, from (\ref{gsqrt3}), we have
\begin{align}
\mu^*(1/\sqrt{3}) = \mu^*(\sqrt{3}) = 1/2.
\label{gstar4}
\end{align}

The global behavior of $\mu^*(K)$ is indicated in Fig. \ref{Fig5}.   In order to reflect the symmetry under $K \leftrightarrow 1/K$ in (\ref{gstarsymmetry}), we plot $K$ on a logarithmic scale.   The red dots and lines indicate the known data in Eqs. \ref{gstar1}-\ref{gstar4}, and the curve is an interpolation.   Of course, the detailed behavior is unknown.   For instance, there could be points where the slope of $\mu^*(K)$ is discontinuous.

We next consider the behavior away from the critical point.   In the AG model, if the symmetry between the two channels is broken by $t_1 - t_2 = \delta t$, then the system flows at low energy to the AN (NA) fixed point for $\delta t>0$ ($\delta t<0$).   At zero temperature, we therefore expect a step function behavior of the conductance: $G_{11}(\delta t) = 2  \theta(\delta t)e^2/h$, $G_{22}(\delta t) = 2 (1-\theta(\delta t))e^2/h$.  This behavior is very similar to the pinch-off transition of the helical point contact.

TK showed that at finite temperature the step is rounded and exhibits a scaling behavior in the limit $T, \delta t \rightarrow 0$:
\begin{align}
G_{11}(T,\delta t) &= (2 K e^2/h) {\cal G}_K(c \delta t/T^{\alpha_K}) \\
G_{22}(T,\delta t) &= (2 K e^2/h) {\cal G}_K(-c \delta t/T^{\alpha_K}).
\label{scalingfun}
\end{align}
Here, $\alpha_K$ is a critical exponent that is determined by the dimension of the leading relevant operator at the critical fixed point and $c$ is a non-universal constant.   ${\cal G}_K(X)$ is a universal crossover scaling function characterizing the flow from the unstable critical point to the stable fixed points, satisfying
\begin{align}
&{\cal G}_K(X=0) =  \mu^*(K) \\
&{\cal G}_K(X \rightarrow \infty) = 1\\
&{\cal G}_K(X \rightarrow \infty)=  0.
\end{align} 

TK computed $\alpha_K$ and ${\cal G}_K(X)$ in the perturbatively accessible regimes.
Here we collect those results.   For $K=1-\epsilon$, 
\begin{equation}
\alpha_{1-\epsilon} = \epsilon^2/2;
\end{equation}
The limiting form of the scaling function is
\begin{equation}
{\cal G}_{1}(X) = \frac{1}{2}\left(1+\frac{X}{\sqrt{1+X^2}}\right).
\label{g1function}
\end{equation}
For $K=1/2 + \epsilon$,
\begin{equation}
\alpha_{1/2+\epsilon} = 4\epsilon
\end{equation}
The limiting form of the scaling function is
\begin{equation}
{\cal G}_{1/2}(X) = \theta(X)\frac{X}{1+X}
\label{g1/2scaling1}
\end{equation}
The singularity near $X=0$ is rounded for finite $\epsilon$.  For $|X|\ll 1$ it is given by
\begin{equation}
{\cal G}_{1/2+\epsilon}(X\ll 1) = \frac{X}{1-e^{-X/(2\pi^2\epsilon)}}.
\label{g1/2scaling2}
\end{equation}
It can be observed that (\ref{g1/2scaling1}) and (\ref{g1/2scaling2}) match for $\epsilon \ll X \ll 1$.  Moreover, (\ref{g1/2scaling2}) and (\ref{gstar1}) agree.

The results for $K=2-\epsilon$ follow from (\ref{gstarsymmetry}).  Note that for $K=1$, the physical conductance is the same as the Kubo conductance, so no translation is necessary for ${\cal G}_1$ in (\ref{g1function})  For $K=1/2$, the TK result requires translation using (\ref{kubotophysicalag}).  Eq. \ref{g1/2scaling1} retains the same form provided we replace $X$ by $X/2$, which changes the non-universal constant $c$ in (\ref{scalingfun}).  The exponent in (\ref{g1/2scaling2}) then differs from TK by a factor of two.   We refer the reader to Ref. \onlinecite{teokane09} for plots of these scaling function. 

\section{Discussion}
\label{section6}

In this paper we have established the equivalence between the Affleck Giuliano model of a two channel Luttinger liquid - topological superconductor junction and the Teo Kane model of a helical point contact.   Both models exhibit a series of phases that are stable, or unstable, depending on the values of the Luttinger interaction parameters.   These phases are identified with the phases of a single impurity in a spinful Luttinger liquid: the charge insulator - spin insulator (II), the charge conductor - spin conductor (CC), as well as the mixed IC and CI phases.   In the TK model, the pinched off limit corresponds to II, while the open limit corresponds to CC.   In the AG model, the limit in which the leads are decoupled from the topological superconductor (leading to normal reflection in both leads, NN) corresponds to the IC phase.    In addition, both models exhibit critical fixed points, which are neither perfectly transmitting or perfectly reflecting, and can be identified with the intermediate fixed points found in the spinfull Luttinger liquid.

The CC, II, IC and CI phases are related by a web of duality transformations.   The TK and AG model are related by the duality that takes II to IC.  When the Luttinger parameter $K=1$, this duality relates two inequivalent free fermion representations of the same problem.   As explained in Section \ref{freefermionsec}, these inequivalent representations are related by the triality of $SO(8)$.

Using the transformation that relates the two models, all predictions for the TK model can be translated to the AG model, and vice versa.   In particular, this analysis allowed us in Section \ref{section5} to predict the global behavior of the critical conductance $G^*(K)$ in the AG model.

There remain a number of open problems for further study.  Foremost among these is to develop a more comprehensive theory of the intermediate fixed points using boundary conformal field theory.   It is generally expected that the fixed points in a quantum impurity problem should be characterized by the set of allowed conformally invariant boundary conditions.   While this is certainly the case for the simple fixed points in our theory, the intermediate fixed points seem to defy this simple classification, and analysis of them has only been possible in certain perturbative limits.   A more general analysis is complicated by the fact that for continuous values of the Luttinger parameter $K$, the conformal field theory describing the leads is not rational.    

In the absence of a general classification, perhaps some progress is possible for specific values of $K$ for which the theory is rational.   For a different regime of the spinful Luttinger liquid problem (with specific values of $K_\rho$ and $K_\sigma$), Yi and Kane\cite{yikane98} mapped the intermediate fixed point to the non-Fermi liquid fixed point of the 3-channel Kondo problem, described by a $SU(2)_3$ Wess-Zumino-Witten theory\cite{affleck91a,affleck91b}.   Perhaps analysis is possible for other values of $K$, corresponding to rational CFT's.
It would also be worthwile to study the equivalent models numerically.

It may also be of interest to consider generalizations.   TK considered a model in which spin conservation could be violated (while preserving time reversal).   That could further be coupled to a superconductor to allow charge conservation to be violated.   Likewise, the AG model can be generalized to include any number of Luttinger liquid leads coupled to the Majorana mode of a topological superconductor.   In their analysis for weak interactions, TK found an additional intermediate fixed point when spin conservation is violated in which an incident electron is transmitted to any of the other three leads with equal probablility 1/3.   It seems likely that this critical point is related to the symmetric critical point of a three lead AG model.   Likewise, when charge conservation is violated, in the TK model, there is likely a non-trivial mapping to the four lead AG model.

\acknowledgments

We thank Paul Fendley for interesting discussions at the commencement of this work and Andrea Nava for insightful discussions about the renormalization group approach for weak interactions.
C.L.K. was supported by a Simons Investigator Grant by the Simons Foundation. I.A. was supported by NSERC Discovery grant 04033-2016. 
D.G. gratefully thanks Steward Blusson Quantum Matter Institute at the University of 
British Columbia for the kind hospitality during the completion of this work.

%

\appendix

\section{Conductance Identity}
\label{appendixa}

In this appendix we prove Eq. (\ref{gxxgyy}), which relates the conductances $G^\rho_{XX}(K)$ and $G^\sigma_{YY}(K)$.  Using the Kubo formula, these conductances can be expressed as correlation functions,
\begin{equation}
G^\alpha_{IJ} = \lim_{\omega\rightarrow 0} \frac{1}{i\omega}\Pi_{IJ}^\alpha(\omega)
\end{equation}
where using (\ref{kuboformula},\ref{irhox},\ref{isigmay}) the retarded correlation functions are given by
\begin{align}
\Pi^\rho_{XX}(t) &=  (K/\pi)^2 \theta(t) \langle [\partial_x\varphi_{\rho-}(t),\partial_x\varphi_{\rho-}(0)]\rangle \label{pirhoxx} \\
\Pi^\sigma_{YY}(t) &=  (\pi K)^{-2} \theta(t) \langle [\partial_x\theta_{\rho-}(t),\partial_x\theta_{\rho-}(0)]\rangle.
\label{pirhoyy}
\end{align}
$\varphi_{\rho-}$ and $\theta_{\rho-}$ obey the commutation relation (\ref{thetaalphacommute}) and have bare Hamiltonian ${\cal H}^0_{\rho-}$ given by (\ref{hrho0}).   Here we set the velocity $v=1$.  The operators in (\ref{pirhoxx},\ref{pirhoyy}) are evaluated at a position $x>0$, away from the junction at $x=0$.

It is useful to transform to a set of chiral currents that diagonalize ${\cal H}^0_{\rho-}$, given by
\begin{align}
J_R &= (\partial_x\varphi_{\rho-} + \partial_x\theta_{\rho-}/K)/(2\pi) \label{chiralfields1}\\
J_L &= (\partial_x\varphi_{\rho-} - \partial_x\theta_{\rho-}/K)/(2\pi).
\label{chiralfields2}
\end{align}
These satisfy
\begin{equation}
[J_R(x),J_R(x')] = -[J_L(x),J_L(x')] =  \frac{i}{2\pi K} \delta'(x-x')
\end{equation}
and $[J_R(x),J_L(x')] = 0$.
In terms of these variables,
\begin{equation}
{\cal H}^0_{\rho-} = \pi v K \left[ J_R^2 + J_L^2 \right].
\end{equation}
Thus, the in and out moving chiral modes are decoupled away from the junction at $x=0$.

Now consider 
\begin{equation}
\Pi^0(t) = K^{-1} \Pi^\rho_{XX}(t) + K \Pi^\sigma_{YY}(t)
\end{equation}
Using, (\ref{chiralfields1},\ref{chiralfields2}), this may be written as
\begin{equation}
\Pi^0(t) = 2K \theta(t) \langle [J_R(t), J_R(0)] + [J_L(t), J_L(0)] \rangle.
\label{pi0eq}
\end{equation}
Note that by design, the cross terms cancel in (\ref{pi0eq}).
When evaluated at a position $x$ away from the junction, the correlator of two right moving (or two left moving) currents will be independent of the Hamiltonian at junction because the left (right) currents at $x$ will be out of causal contact with $x=0$.  

$\Pi^0(t)$ can be straightforwardly evaluated, but it is even simpler to see that
\begin{equation}
G_0 = K^{-1} G^\rho_{XX} + K G^\sigma_{YY}
\end{equation}
must independent of the tunneling at the barrier.   Therefore it can be trivially evaluated at either the II or CC fixed point.  At CC, $G^\rho_{XX} = 2K e^2/h$ and $G^\sigma_{YY} = 0$.   Therefore,
\begin{equation}
G_0 = 2 e^2/h.
\end{equation}

\section{Dual coupling to the Majorana mode in the AA limit of the AG model}
\label{k2e}

A direct derivation of the dual AA action in the AG model, (\ref{s0aa},\ref{vaa})  of the main text,
can be performed within a lattice fermion description of the system, which eventually 
yields  (\ref{s0aa},\ref{vaa})  in the continuum limit \cite{giuliano19}. As  lattice 
fermion Hamiltonian for the AG model we use  $H_{\rm Lat}$,  given  by

\begin{eqnarray}
 H_{\rm Lat}  &=& \sum_{a = 1}^2 \sum_{ j = 1}^{\infty}  \{ - J [ c_{j , a}^\dagger c_{j + 1 , a}   + c_{ j + 1 , a}^\dagger c_{j , a} ] 
 - \mu c_{ j , a}^\dagger c_{j , a} \} \nonumber \\
 &-&    \sum_{a = 1}^2 t_{a } \gamma  \: \{ c_{a , 1} - c_{a , 1}^\dagger \} + H_{\rm Int} 
   \:\:\:\: , 
   \label{bi.3}
\end{eqnarray}
\noindent
with $J$ being the lattice hopping strength and $\mu$ being the chemical potential. 

The first term at the second line of (\ref{bi.3}) is the lattice version of the coupling to the Majorana mode. 
$H_{\rm Int}$ is the lattice bulk interaction Hamiltonian. Its explicit expression is not relevant for 
the following derivation, provided that, in the continuum limit, one recovers the 
Luttinger liquid Hamiltonian in (\ref{lut_AG1},\ref{lut_AG2}) (details about $H_{\rm Int}$ are 
provided in Ref. \onlinecite{affleck13}). The standard pathway from (\ref{bi.3}) to the Luttinger liquid 
Hamiltonian goes through  retaining only low-energy, long-wavelength fermionic modes of 
the lattice fermion operators by expanding them as  
$c_{j , a} \sim  \{  e^{ i k_F j}  \Psi_{+ , a} ( x ) + e^{ - i k_F j }  \Psi_{- , a} ( x ) \}$,  
with the Fermi momentum $\pm k_F =  {\rm arccos} \left( - \frac{\mu}{2 J} \right)$, and by therefore 
employing the bosonization formulas of the main text for the continuum fermion operators.

As a next step, we now introduce   lattice real fermion operators  $\{ \xi_{j , a} , \eta_{j , a } \}$, 
such that  $c_{j , a} =  \xi_{j , a } + i \eta_{j , a}$.  Also, for the 
sake of simplicity, we make the assumption of symmetric couplings to the Majorana mode 
in the second line of (\ref{bi.3}), that is   $t_1 = t_2 = t$. As a result, the 
corresponding term in (\ref{bi.3}) is given by  $ - 8 i t \sum_{a = 1,2} \: \gamma \eta_{1,a} = - 8 i t \sqrt{2} \gamma \eta_+$,
with $\eta_+ = \frac{1}{\sqrt{2}} ( \eta_1 + \eta_2)$. In the large $t$-limit, we therefore see that 
$\eta_+$ is ``locked together'' with $\gamma$ into a state annihilated by the Dirac operator 
$\gamma + i \eta_+$.

The real fermion  operators $\xi_{1,a}$ appear in the free Hamiltonian at the 
first line of  (\ref{bi.3}). Putting the corresponding contributions all together, we 
define $H'$ given by   

\begin{equation}
H' = - 2 i \: \sum_{a=1,2} \: \{ \xi_{2 ,a} \eta_{1,a} + \xi_{1,a} \eta_{2,a} \} 
- 2 i \mu \:  \sum_{a=1,2} \:  \xi_{1 ,a} \eta_{1,a}
\:\:\:\: . 
\label{ls.3}
\end{equation}
\noindent
$H'$ can be regarded as a special case of a general  Hamiltonian $\tilde{H}'$, given by 

\begin{eqnarray}
\tilde{H}' &=& - 2 iJ_\alpha  \: \sum_{a=1,2} \:   \xi_{2 ,a} \eta_{1,a}  - 2 iJ_\beta  \: \sum_{a=1,2} \: \xi_{1,a} \eta_{2,a}  \nonumber \\
&-& 2 i \mu_\alpha \: \sum_{a=1,2} \:  \xi_{1 ,a} \eta_{1,a}
\:\:\:\: ,
\label{ls.4}
\end{eqnarray}
\noindent
with $J_\alpha = J_\beta = J$ , $\mu_\alpha = \mu$. 

To stabilize the AA fixed point, one therefore fine-tunes  the 
parameters in $\tilde{H}'$ so that $J_\alpha = \mu_\alpha = 0$, $J_\beta = J \neq 0$. As a result, 
the first line of (\ref{bi.3}) becomes  

\begin{eqnarray}
\tilde{H} &=& \sum_{a = 1,2}\sum_{j = 2}^{\infty } \:  \{  - J  [  c_{j , a}^\dagger c_{a+1 , a} + c_{j+1,a}^\dagger c_{j , a} ]  \nonumber \\
&-& \mu   c_{j , a}^\dagger c_{j , a }  \} - J  \sum_{a=1,2} \xi_{1,a} \: \{ c_{2,a} - c_{2,a}^\dagger \} 
\:\:\:\: . 
\label{ls.4bis}
\end{eqnarray}
\noindent
$\tilde{H}$ in (\ref{ls.4bis}) is the sum of two Hamiltonians for a single chain coupled to a Majorana mode at its endpoint. Therefore,  
the fixed point corresponds to perfect Andreev reflection in both channels, that is to having (in terms of the continuum fields), 
$\Psi_{+,a} ( 0 ) = \Psi_{-,a}^\dagger ( 0 )$, for $a=1,2$. This result is not affected by turning on the bulk interaction, 
provided the Luttinger parameter $K > 1/2$ \cite{fidkowski12,affleck13}. 
 
 Turning on the  terms we set to zero before,   we obtain the corresponding  boundary interaction, given by  

\begin{eqnarray}
\Delta \tilde{H}' &=& 
- i  \sqrt{2} \sum_{a=1,2} \: \{ J_a \xi_{2,a} + \mu_a  \xi_{1,a} \} \eta_+   \nonumber \\
&+& i   \sqrt{2}  \sum_{a=1,2} \: \{ (-1)^a [ J_a  \xi_{2,a} + \mu_a  \xi_{1,a} ] \} \eta_-
\:\:\:\: .
\label{ls.6}
\end{eqnarray}
\noindent
Due to the locking between $\eta_+$ and $\gamma$, any action of $\eta_+$ alone would take the system
to a higher energy state, which we forbid, in the large-$t$ limit. Therefore, in the following we 
drop the term $\propto \eta_+$ from the right-hand side of (\ref{ls.6}). Expanding the lattice fermion 
operators in terms of the continuum fields, taking into account the AA boundary conditions 
and resorting to bosonization framework, one eventually obtains

\begin{equation}
\Delta \tilde{H}'  =    i \sum_{a = 1,2}  \eta_- v_a \Gamma_a \cos   \theta_a  
\:\:\:\: ,
\label{are,x13}
\end{equation}
\noindent
that is, $V_{AA}$ in (\ref{vaa}) of the main text, with $\eta_-$ being the ``emerging'' Majorana mode 
at the AA fixed point.

\section{Renormalization group analysis for weak interactions}
\label{rg_appe}

In this appendix we explicitly derive the renormalization group flow equations for ${\cal S}^{AG}$ for small values of 
the bulk interaction strength $V$ (that is, for $| V / ( 2 \pi v ) | = | \epsilon | (= | 1-K| ) \ll 1$), 
up to order $\epsilon^2$. While we derive the equations for a generic ${\cal S}$-matrix, we show that, 
when ${\cal S} = {\cal S}^{AG}$ and  ${\rm det} [{\cal S}^{AG} ] = -1$, the term linear in $\epsilon$ vanishes and 
the first nonzero contribution to the renormalization group equations appears to order $\epsilon^2$. This is anologous to 
what happens to the renormalization group equations for ${\cal S}^{TK}$ at small values of $\epsilon$, as discussed in 
Ref. \onlinecite{teokane09} (what is basically expected, due to the correspondence between the two models). 

To perform our derivation, we picture the AG model as a   
quantum point contact between $N=4$ interacting leads, so to employ the general framework developed in Refs. \onlinecite{matveev93,yue94,teokane09}. 
Let $\psi_{a,p}$, with $a = 1,2,3,4$ and $p = \pm 1$, the in- and out-fermionic field operators in each lead. 
When deriving the equations for the AG model,  
we identify  $\psi_{1,p} , \psi_{2,p}$ with respectively  $\Psi_{1,p}$ and $\Psi_{2,p}$ of the 
AG model, and $\psi_{3,p} , \psi_{4,p}$ with respectively $\Psi_{1,p}^\dagger , \Psi_{2,p}^\dagger$. 
To consistently realize the identification,  
we employ a generalization of TK bulk interaction  Hamiltonian \cite{teokane09}, with an interaction strength that has  
off-diagonal components in the lead index. Specifically, we use $H_{\rm Int}$ given by  
\begin{eqnarray}
H_{\rm Int}  &=&  ( 2 \pi v \epsilon )
\sum_{a, b  } \: v_{ab} \nonumber \\
&\times& \int_0^\infty  \: d x \: \psi_{a , + }^\dagger ( x ) \psi_{a , + } ( x ) \psi_{b,- }^\dagger (x ) 
\psi_{b,-  } ( x ) 
\:\:\:\: , 
\label{ns.1}
\end{eqnarray}
\noindent
with $v_{a,b} = \delta_{a,b} - \delta_{a,b+2}$, and $a,b$
understood to be defined modulo 4. The fields $\psi_{a,+}$ and $\psi_{a,-}$ are related to each other by 
the ${\cal S}$-matrix via 
\begin{equation}
 \psi_{a,+} = \sum_{b } \: {\cal S}_{ab} \psi_{b,-} 
 \:\:\:\: .
 \label{ns.x2}
\end{equation}
\noindent
(\ref{ns.x2}) allows for relating the ${\cal S}$-matrix elements to the single-fermion 
Green's functions involving a $+$ and a $-$ field. In particular, resorting to the imaginary time 
formalism and using $\psi_{a,p} ( x , \tau )$ to denote the field $\psi_{a , p } ( x )$ at imaginary
time $\tau$, one obtains \cite{teokane09}
\begin{eqnarray}
&& \: \left[ \begin{array}{cc}  g_{a,b}^{(+,+)}  ( x , \tau ; x' , \tau' ) 
&       g_{a,b}^{(+,-)}  ( x , \tau ; x' , \tau' )     \\    g_{a,b}^{(- ,+)}  ( x , \tau ; x' , \tau' ) 
& g_{a,b}^{(-,-)}  ( x , \tau ; x' , \tau' )  \end{array} \right] 
 \equiv \nonumber \\  
 && = \left[ \begin{array}{cc}  \delta_{ab} \;  g (x , \tau ; x' , \tau') &  
 {\cal S}_{ab} \;   g (x , \tau ; - x' , \tau')  \\ {\cal S}^\dagger_{ab} \;   g (-x , \tau ;  x' , \tau') 
 &  \delta_{ab} \;  g (-x , \tau ; -x' , \tau') 
 \end{array} \right] 
 \;\;\;\; , 
 \label{adns.1}
\end{eqnarray}
\noindent
with $g_{a,b}^{(p,p')} ( x , \tau ; x' , \tau' ) = - i  \langle {\bf T}_\tau \psi_{a ,p } ( x , \tau ) \psi_{b , p' }^\dagger 
 (x' , \tau' ) \rangle$, $g ( x , \tau ; x' , \tau' ) = \frac{1}{2 \pi i } \: \left( \frac{1}{v \tau + i x - v \tau' - i x' } \right)$, 
and with  ${\bf T}_\tau$ being the imaginary time ordering operator. 
According to (\ref{adns.1}),
we compute  the correction to ${\cal S}$ at a given order in $\epsilon$ by just looking at 
the corresponding correction to   $g_{a,b}^{(+,-)}  ( x , \tau ; x' , \tau' )$ in 
 (\ref{adns.1}). In particular, using the interaction Hamiltonian in 
 (\ref{ns.1}) and denoting with $\delta^{(1)} g_{a,b}^{(+,-)} ( x , \tau ; x' , \tau')$ the 
correction to order $\epsilon$ to $g_{a,b}^{(+,-)} ( x , \tau ; x' , \tau' )$, one finds 
\begin{eqnarray}
 &&  \delta^{(1)}  g_{a,b}^{(+,-)}( x , \tau ; - x' , \tau' ) 
 =  (2 \pi v \epsilon ) \: \sum_{d,d'  }  \: v_{dd'}  \sum_{c, c' , c''  } 
  S_{ac} S_{dc'}^* S_{d c''}   \nonumber \\
 && \:\times \int_{-\infty}^\infty \: d \tau_1 \: \int_0^\infty \: d y_1  \langle {\bf T}_\tau \psi_{c , +} ( x , \tau ) \psi_{c' , + }^\dagger ( y_1 ,  \tau_1 ) \psi_{c'' , +} ( y_1 , \tau_1 ) 
\nonumber \\
&&  \times 
 \psi_{d' , -}^\dagger  ( y_1  , \tau_1 ) \psi_{d' , -} (  y_1  , \tau_1 ) \psi_{b , -}^\dagger ( x' ,  \tau' ) \rangle 
 \:\:\:\: . 
 \label{monoz.5}
\end{eqnarray}
\noindent
Applying Wick's theorem to the last line of  (\ref{monoz.5}) and taking into account  (\ref{adns.1}) as well 
as the unitarity of ${\cal S}$, one eventually recasts  (\ref{monoz.5}) in the form 
\begin{eqnarray}
 &&  \delta^{(1)}  g_{a,b}^{(+,-)}( x , \tau ; - x' , \tau' ) 
 =  - ( 2 \pi v \epsilon ) v_{ab} {\cal S}_{ab} \int_{-\infty}^\infty \: d \tau_1 \: \int_0^\infty \: d y_1  \nonumber \\
 && \times  
 g ( x , \tau ; y_1 , \tau_1 ) g ( y_1 , \tau_1 ; - y_1 , \tau_1 ) 
 g ( - y_1 , \tau_1 ; - x' , \tau' ) \nonumber \\
 && + ( 2 \pi v \epsilon ) \: \sum_{c , d }  v_{cd} {\cal S}_{ad} {\cal S}^*_{cd} {\cal S}_{cb} 
  \int_{-\infty}^\infty \: d \tau_1 \: \int_0^\infty \: d y_1 \nonumber \\
  && \times g ( x , \tau ; - y_1 , \tau_1 ) g ( y_1 , \tau_1 ; - y_1 , \tau_1 ) 
 g ( y_1 , \tau_1 ; -x' , \tau' ) 
  \:\:\:\:\: . 
 \label{monoz.5b}
 \end{eqnarray}
\noindent
In Fig. \ref{processes_first} we diagrammatically draw the physical processes contributing the 
right-hand side of  (\ref{monoz.5}).  To ease reading the graphs, 
we employ a full red line to denote an in-particle ($p=-$) and a full blue line to denote 
an out-particle ($p=+$).   
We represent the interaction  as a dashed line connecting the densities in the in and out channels.  A scattering event 
at the junction, connecting in and out particles, corresponding to a single insertion of an ${\cal S}$-matrix element, is denoted with a 
full black dot.   Due to the off diagonal form of our interaction, only diagrams with one ((Fig. {\ref{processes_first}(a)) or three (Fig. {\ref{processes_first}(b)) ${\cal S}$-matrix insertions contribute to the renormalization of ${\cal S}$.

To regularize the logarithmic divergences, one resorts to the standard renormalization group 
approach \cite{matveev93,yue94,teokane09}, by introducing  the running parameter $l = \ln \left( \frac{D_0}{D}\right)$, with 
$D$ being the running energy scale and $D_0$ being a high energy, reference cutoff.
Taking into account the contributions represented in the diagrams of Fig. \ref{processes_first},
one eventually obtains the renormalization group equations for   ${\cal S}_{ab}$ to first order 
in $\epsilon$, given by 
\begin{equation}
\frac{ d {\cal S}_{ab} }{d l }= 
  \frac{\epsilon}{2} \{ v_{ab}  {\cal S}_{ab}-  \sum_{c,d} \: v_{cd} 
  {\cal S}_{ad} {\cal S}_{cd}^* {\cal S}_{cb} \} + \ldots 
  \:\:\:\: , 
  \label{froc.6}
 \end{equation}
 \noindent
with the ellipses corresponding to higher-order contributions, which we are 
going to compute in the following.  
 
    \begin{figure}
 \center
\includegraphics*[width=1. \linewidth]{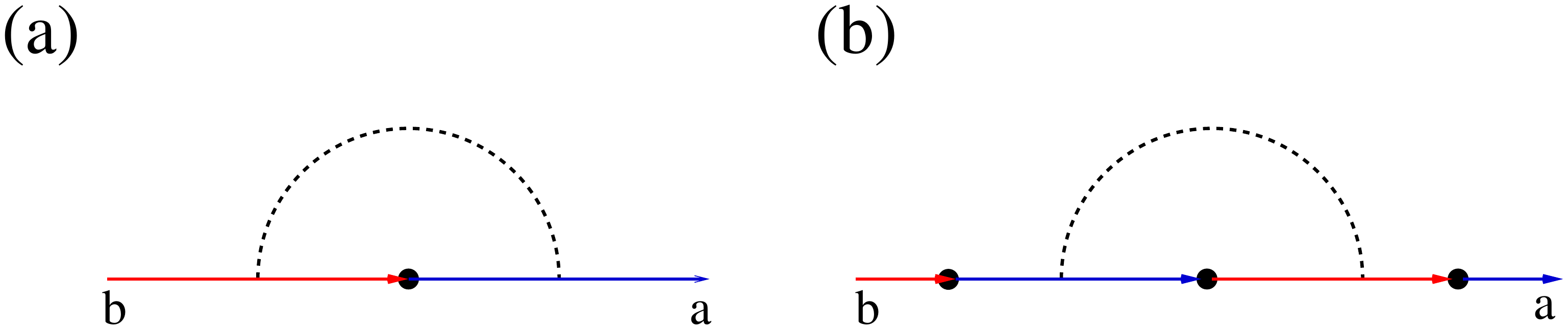}
\caption{Diagrammatic representation of the physical processes that, to order $\epsilon$, correct ${\cal S}$ 
by logarithmically diverging contributions. In the figure, a red full line represents an in-particle, a 
blue full line an out-particle, the dashed black line corresponds to the bulk interaction, and the black 
dots denote insertions of ${\cal S}$-matrix elements.  In particular, in the diagram (a) the bulk interaction converts a red line into a red line, 
as well as a blue line into a blue line, thus providing a correction that is first-order in the ${\cal S}_{ab}$'s. In 
the diagram (b), instead, the bulk interaction converts a red line into a blue line and vice versa, thus leading to
a correction that is third-order in the ${\cal S}_{ab}$'s.   } 
\label{processes_first}
\end{figure}

In the quantum point contact between four 
helical states and without inter-wire interaction discussed in Ref. \onlinecite{teokane09}, 
that is, for ${\cal S} = {\cal S}^{TK}$, one has  $v_{ab} = \delta_{ab}$ and  ${\cal S}^{TK}_{aa} = 0$. Apparently, this implies that 
the right-hand side of  (\ref{froc.6}) is =0 $\forall a,b$.  In the AG model, that is, for 
${\cal S} = {\cal S}^{AG}$, the key point is whether ${\rm det} [ {\cal S}^{AG} ] = -1$, as it happens 
when the Majorana zero mode at the junction is ``built in'' the scattering boundary conditions (as at the e.g. 
AN and NA fixed points of the AG model), or not (as it happens at the NN and AA fixed points). 
If ${\rm det} [{\cal S}^{AG}] = -1$, then ${\cal S}^{AG}$ can be expressed using the 
completely general parametrization in  (\ref{sss}).  As a result,  plugging the corresponding matrix elements in the right-hand 
side of  (\ref{froc.6}), one finds that, exactly as it happens in the TK model, the renormalization 
group equations for ${\cal S}^{AG}$ are zero, to order $\epsilon$ (which corresponds to a perfect cancellation 
between the amplitudes corresponding to the diagrams in Fig. \ref{processes_first}(a) and 
\ref{processes_first}(b)). So, to get a finite result, one has to derive the renormalization 
group equations to order $\epsilon^2$. 

At variance, when    ${\rm det} [ {\cal S}^{AG} ] = 1$,    ${\cal S}^{AG}$ is 
parametrized by (\ref{sss}) by simply swapping with each other rows 2 and 4 in   
(which reverses the sign of ${\rm det} [{\cal S}^{AG} ] $).
 Plugging again  the matrix elements of ${\cal S}^{AG}$  
 into  (\ref{froc.6}), one finds that the right-hand side 
of the equation is no longer equal to 0. Indeed, to order $\epsilon$ one 
obtains the  differential equations for the transmission coefficients 
$T_1 = | t_1|^2$ and $T_2 = | t_2 |^2$ which, as it appears from 
(\ref{c.1},\ref{c.2},\ref{c.3},\ref{c.4}), fully characterize the transport properties of the junction.
Specifically, the equations for $T_1,T_2$ are given by

\begin{eqnarray}
\frac{ d T_1 }{d l }&=& - \epsilon T_1 ( 1 - T_1 )   \nonumber \\
\frac{d T_2 }{d l }&=& - \epsilon T_2 ( 1 -  T_2 ) 
\:\:\:\: , 
\label{ck.bx1}
\end{eqnarray}
\noindent
which is the appropriate generalization of the main result of Refs. \onlinecite{matveev93,yue94}.

    \begin{figure}
 \center
\includegraphics*[width=1. \linewidth]{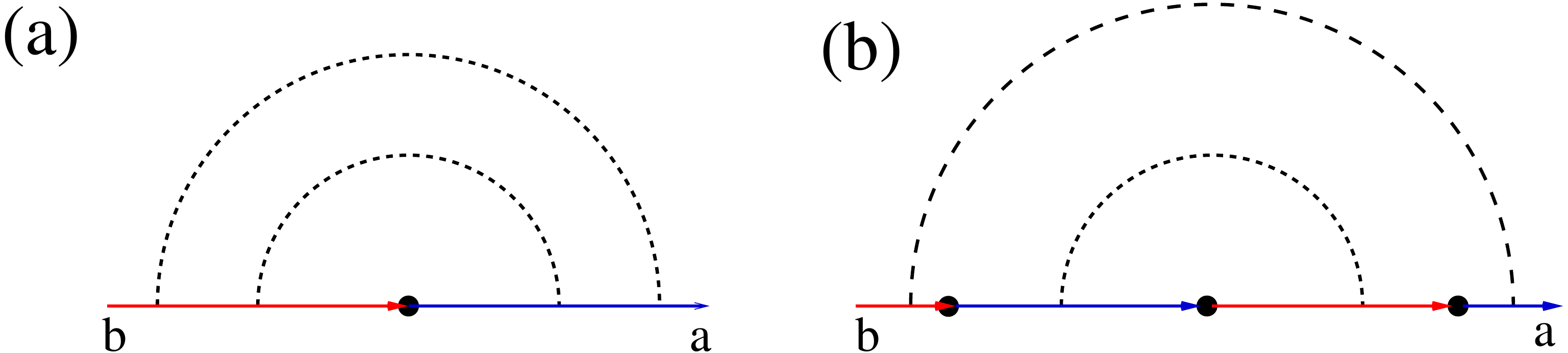}
\caption{First pair of diagrams that, to order $\epsilon^2$, cancel with each other, as a consequence of the cancellation 
between diagrams of order $\epsilon$ in Fig. \ref{processes_first}.}  
\label{processes_second_c1}
\end{figure}
    \begin{figure}
 \center
\includegraphics*[width=1.  \linewidth]{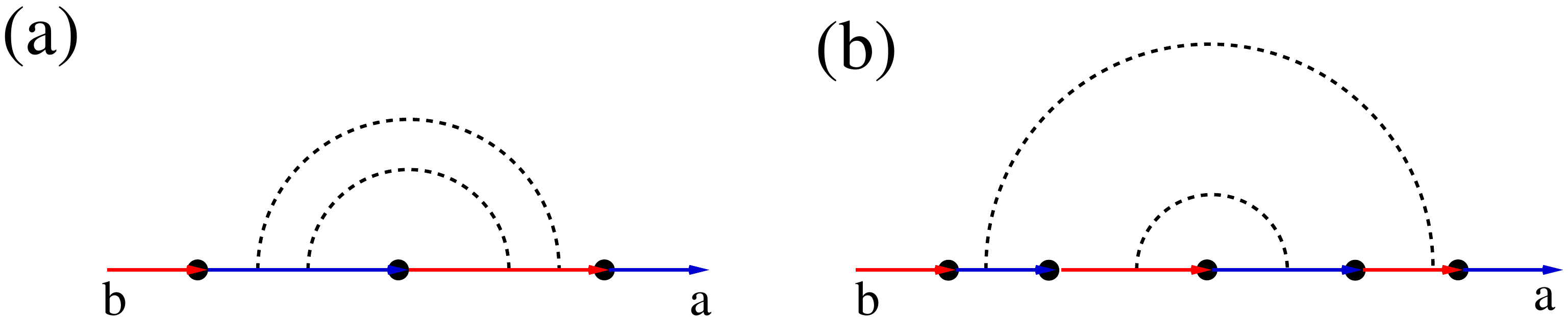}
\caption{Second pair of diagrams that, to order $\epsilon^2$, cancel with each other, as a consequence of the cancellation 
between diagrams of order $\epsilon$ in Fig. \ref{processes_first}.} 
\label{processes_second_c2}
\end{figure}
 
Deriving the renormalization 
group equations for ${\cal S} $  to   order $\epsilon^2$, is straightforward, though 
tedious, due to the large number of independent diagrams all of which, in principle, can potentially contribute 
logarithmically diverging corrections to the ${\cal S}$ matrix elements. However, as it happens in the   analogous calculation 
performed within the TK model \cite{teokane09}, many of the diagrams either provide finite corrections not contributing 
to the renormalization of ${\cal S}$, or just come with opposite sign and cancel with each other 
(in this respect, they behave in analogy to the diagrams in Fig. 9(b) and 9(c) of Ref. \onlinecite{teokane09}), 
or cancel with each other, as a consequence of the fact that the diagrams 
to order $\epsilon$ in Fig. \ref{processes_first}(a) and \ref{processes_first}(b) cancel with each other.
In Figs. \ref{processes_second_c1} and \ref{processes_second_c2}, we draw pairs of 
diagrams to order $\epsilon^2$ which, when summed to each other, contain a factor proportional to the sum of the diagrams in 
Fig. \ref{processes_first}(a) and \ref{processes_first}(b). In fact, the cancellation is apparent, once one 
compares the diagrams in Fig. \ref{processes_first} with the ``inner'' part of the diagrams in 
Figs. \ref{processes_second_c1} and \ref{processes_second_c2} (that is, the portion of each diagram consisting of the inner 
dashed line together with all the full lines connected to it). 

Taking into account the cancellations encoded in the diagrams in Figs. \ref{processes_second_c1} and \ref{processes_second_c2}, 
 one finds that the only independent diagrams  that, to order $\epsilon^2$, provide nonzero contributions to the renormalization group equations are 
the ones drawn in Fig. \ref{processes_second}. Putting all together the contributions of the diagrams in 
Fig. \ref{processes_second}, we eventually obtain the renormalization group equations for 
the ${\cal S}_{ab}$'s to order $\epsilon^2$. Assuming that the
terms of order $\epsilon$ are = 0, these are given by 

\begin{eqnarray}
\frac{d {\cal S}_{ab}}{ d l } &=& \frac{\epsilon^2}{4} \: \{     \sum_{cd}v_{ad} v_{cb} {\cal S}_{ab} | {\cal S}_{cd} |^2
\nonumber \\
&-&  \sum_{c,d,g,h}v_{ch} v_{gd}  {\cal S}_{ad}  {\cal S}^*_{cd} {\cal S}_{cb} |  {\cal S}_{gh}|^2 \} 
\:\:\:\: . 
\label{final}
\end{eqnarray}
\noindent
 (\ref{final}) is the  main result of this appendix.  To check that it consistently generalizes  (3.39) of 
Ref. \onlinecite{teokane09}, one can assume  a purely diagonal (in the channel index) ``bulk'' 
interaction, that is, one can  set $v_{ab} = \delta_{ab}$. As a result,  (\ref{final}) reduces to

\begin{equation}
\frac{d {\cal S}_{ab}}{ d l } = \frac{\epsilon^2}{4} \: \{   {\cal S}_{ab}   | {\cal S}_{ab} |^2 
- \sum_{cd } {\cal S}_{ad}{\cal S}_{cd}^* {\cal S}_{cb} | {\cal S}_{cd}|^2 \}
\:\:\:\: , 
\label{final.2}
\end{equation}
\noindent 
that is,    (3.39) of Ref. \onlinecite{teokane09}.

    \begin{figure}
 \center
\includegraphics*[width=1.\linewidth]{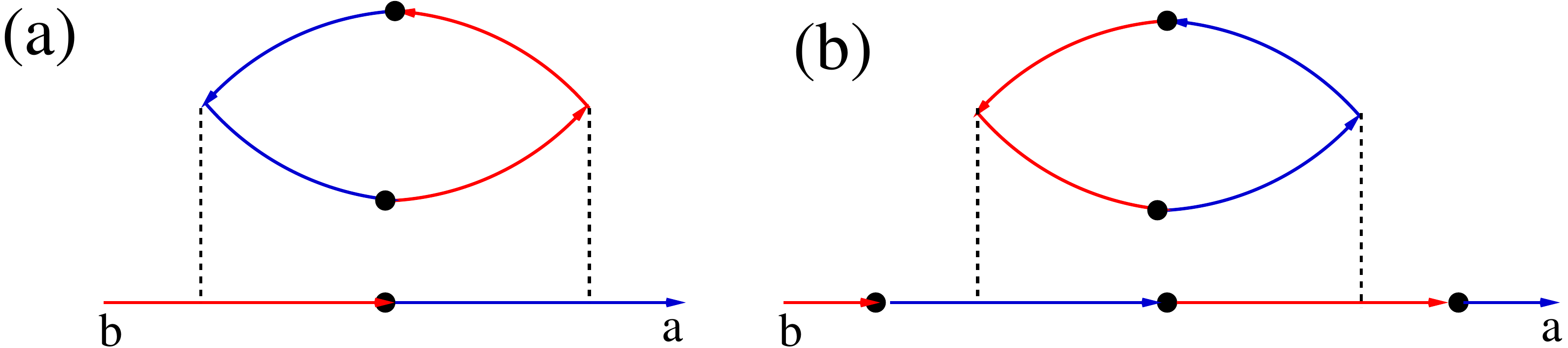}
\caption{Diagrammatic representation of the physical processes that, to order $\epsilon^2$, correct ${\cal S}$ 
by logarithmically diverging contributions. The drawing code is the same as we used for Fig. \ref{processes_first}.
In particular, the diagram (a) provides the term at the right-hand side of  (\ref{final}) that is 
of third order in the ${\cal S}_{ab}$'s, the diagram (b), instead, provides the contribution that 
is of fifth order in the ${\cal S}_{ab}$'s. Remarkably,  the  diagrams in 
(a) and (b)  respectively correspond to the diagrams in Fig. 9(d) and Fig. 9(e) of Ref. \onlinecite{teokane09}, 
generalized to the case of nonzero inter-channel interaction.} 
\label{processes_second}
\end{figure}
\noindent
When ${\cal S} = {\cal S}^{AG}$, using the parametrization  (\ref{sss}),   
one can reexpress the renormalization group equations in  (\ref{final}) in terms of $T_1 =t_1^2$ 
and $T_2 = t_2^2$. As a result, one obtains 
\begin{eqnarray}
 \frac{ d T_1 }{d l } &=& -  \epsilon^2   \:T_1  ( 1 - T_1  )   ( 1 - 2 T_2 )  \nonumber \\
  \frac{ d T_2}{d l } &=& -  \epsilon^2  \: T_2  ( 1 -   T_2 )  ( 1 - 2 T_1 ) 
 \:\:\:\: . 
 \label{final.3}
\end{eqnarray}
That is  (125,126) of the main text. As a consistency check, we note that, if $T_1 = T_2 \equiv T$, 
 (\ref{final.3}) reduces to  (3.41) of Ref. \onlinecite{teokane09}, as  expected.

\smallskip
\section{Kubo conductance vs. physical conductance}
\label{appendixc}

It is well known that the Kubo formula for the conductance does not properly account for the DC conductance measured with Fermi liquid leads\cite{maslov95,safi95}.  
The Kubo conductance
describes the response of an infinite Luttinger liquid at finite frequency, where
the limit $L\rightarrow \infty$ is taken before $\omega\rightarrow 0$. This does not take into
account the contact resistance between the Luttinger liquid
and the electron reservoir where the voltage is defined. An
appropriate model to account for this is to consider a 1D
model for the leads in which the Luttinger parameter $K=1$ for $x>L$.   The physical conductance can therefore be computed using the Kubo formula in a model in which the interactions are turned off for $x>L$.

The relation between the physical conductance $G^{\rm phys}$ computed in this way and the Kubo conductance $G^{\rm Kubo}$ computed with infinite Luttinger liquid leads has been discussed previously\cite{chamon03,oshikawa06,teokane09}.   When the leads have Luttinger parameter $K$, there is effectively an additional contact resistance $R^c= (h/e^2)(K-1)/(2K)$.  Here we simply quote the relevant results.  

For the TK model, with $I= X$ or $Y$, and $\alpha = \rho$ or $\sigma$,
\begin{equation}
\frac{1}{G^{\alpha, {\rm phys}}_{II}(K)} = \frac{1}{G^{\rho, {\rm Kubo}}_{II}(K)} + \frac{h}{e^2}\frac{K_\alpha-1}{2 K_\alpha},
\end{equation}
with $K_\rho = K$, $K_\sigma = 1/K$.
For the AG model, with $A = 1$ or $2$, 
\begin{equation}
\frac{1}{G^{{\rm phys}}_{AA}(K)} = \frac{1}{G^{ {\rm Kubo}}_{AA}(K)} + \frac{h}{e^2}\frac{K-1}{2 K}.
\label{kubotophysicalag}
\end{equation}

\end{document}